\documentstyle [12pt,eqsecnum,aps,amsfonts] {revtex}
\input epsf
\newcommand{\beq}{\begin{equation}}
\newcommand{\eeq}{\end{equation}}
\newcommand{\beqs}{\begin{eqnarray}}
\newcommand{\eeqs}{\end{eqnarray}}

\begin{document}
\tighten
\draft

\baselineskip 6.0mm

\title{Chromatic Polynomials for Lattice Strips with Cyclic Boundary
Conditions}

\author{Shu-Chiuan Chang\thanks{email: shu-chiuan.chang@sunysb.edu}}

\address{
C. N. Yang Institute for Theoretical Physics  \\
State University of New York       \\
Stony Brook, N. Y. 11794-3840  \\
USA}

\maketitle

\vspace{10mm}

\begin{abstract}

The zero-temperature $q$-state Potts model partition function for a lattice
strip of fixed width $L_y$ and arbitrary length $L_x$ has the form
$P(G,q)=\sum_{j=1}^{N_{G,\lambda}}c_{G,j}(\lambda_{G,j})^{L_x}$, and is
equivalent to the chromatic polynomial for this graph. We present exact
zero-temperature partition functions for strips of several lattices with
$(FBC_y,PBC_x)$, i.e., cyclic, boundary conditions. In particular, the
chromatic polynomial of a family of generalized dodecahedra graphs is
calculated.  The coefficient $c_{G,j}$ of degree $d$ in $q$ is
$c^{(d)}=U_{2d}(\frac{\sqrt{q}}{2})$, where $U_n(x)$ is the Chebyshev
polynomial of the second kind.  We also present the chromatic polynomial for
the strip of the square lattice with $(PBC_y,PBC_x)$, i.e., toroidal, boundary
conditions and width $L_y=4$ with the property that each set of four vertical 
vertices forms a tetrahedron.  A number of interesting and novel features of
the continuous accumulation set of the chromatic zeros, ${\cal B}$ are found. 

\end{abstract}

\vspace{16mm}

\pagestyle{empty}
\newpage

\pagestyle{plain}
\pagenumbering{arabic}
\renewcommand{\thefootnote}{\arabic{footnote}}
\setcounter{footnote}{0}

\section{Introduction}

The $q$-state Potts antiferromagnet (AF) \cite{potts,wurev} exhibits nonzero
ground state entropy, $S_0 > 0$ (without frustration) for sufficiently large
$q$ on a given lattice $\Lambda$ or, more generally, on a graph $G$.  This is
equivalent to a ground state degeneracy per site $W > 1$, since $S_0 = k_B \ln
W$.  There is a close connection with graph theory here, since the
zero-temperature partition function of the above-mentioned $q$-state Potts
antiferromagnet on a graph $G$ satisfies

\beq 
Z(G,q,T=0)_{PAF}=P(G,q) \ ,
\label{zp} 
\eeq 
where $P(G,q)$ is the chromatic polynomial expressing the number of ways
of coloring the vertices of the graph $G$ with $q$ colors such that no two
adjacent vertices have the same color (for reviews, see 
\cite{rrev}-\cite{bbook}).  The minimum number of colors necessary for
such a coloring of $G$ is called the chromatic number, $\chi(G)$.  Thus

\beq 
W(\{G\},q) = \lim_{n \to \infty} P(G,q)^{1/n} \ ,
\label{w} 
\eeq 
where $n$ is the number of vertices of $G$ and $\{G\} = \lim_{n \to 
\infty}G$.  Where no confusion will result, we shall sometimes write $G$ rather
than $\{G\}$ for the infinite-length limit of a given type of strip graph. 
At certain special points $q_s$ (typically $q_s=0,1,..,
\chi(G)$), one has the noncommutativity of limits 

\beq 
\lim_{q \to q_s} \lim_{n \to \infty} P(G,q)^{1/n} \ne \lim_{n \to \infty}
\lim_{q \to q_s}P(G,q)^{1/n} \ ,
\label{wnoncom} 
\eeq 
and hence it is necessary to specify the order of the limits in the
definition of $W(\{G\},q_s)$ \cite{w}. Denoting $W_{qn}$ and $W_{nq}$ as
the functions defined by the different order of limits on the left and
right-hand sides of (\ref{wnoncom}), we take $W \equiv W_{qn}$ here; this
has the advantage of removing certain isolated discontinuities that are
present in $W_{nq}$. 

Using the expression for $P(G,q)$, one can generalize $q$ from ${\mathbb
Z}_+$ to ${\mathbb C}$.  The zeros of $P(G,q)$ in the complex $q$ plane
are called chromatic zeros; a subset of these may form an accumulation set
in the $n \to \infty$ limit, denoted ${\cal B}$, which is the continuous
locus of points where $W(\{G\},q)$ is nonanalytic. 
\footnote{\footnotesize{For some families of graphs ${\cal B}$ may be
null, and $W$ may also be nonanalytic at certain discrete points.}} The
maximal region in the complex $q$ plane to which one can analytically
continue the function $W(\{G\},q)$ from physical values where there is
nonzero ground state entropy is denoted $R_1$.  The maximal value of $q$
where ${\cal B}$ intersects the (positive) real axis is labeled
$q_c(\{G\})$.  This point is important since $W(\{G\},q)$ is a real
analytic function from large values of $q$ down to $q_c(\{G\})$. 

\unitlength 1.3mm
%cyclic 55 lattice with L_y=4
\begin{picture}(100,30)
\multiput(0,0)(10,0){5}{\circle*{2}}
\multiput(0,10)(10,0){5}{\circle*{2}}
\multiput(0,20)(10,0){5}{\circle*{2}}
\multiput(0,30)(10,0){5}{\circle*{2}}
\multiput(0,0)(10,0){5}{\line(0,1){30}}
\multiput(0,0)(0,30){2}{\line(1,0){40}}
\multiput(0,10)(10,0){4}{\line(1,1){10}}
\put(-2,-2){\makebox(0,0){13}}
\put(8,-2){\makebox(0,0){14}}
\put(18,-2){\makebox(0,0){15}}
\put(28,-2){\makebox(0,0){16}}
\put(38,-2){\makebox(0,0){13}}
\put(-2,12){\makebox(0,0){9}}
\put(8,12){\makebox(0,0){10}}
\put(18,12){\makebox(0,0){11}}
\put(28,12){\makebox(0,0){12}}
\put(38,12){\makebox(0,0){9}}
\put(-2,22){\makebox(0,0){5}}
\put(8,22){\makebox(0,0){6}} 
\put(18,22){\makebox(0,0){7}} 
\put(28,22){\makebox(0,0){8}} 
\put(38,22){\makebox(0,0){5}}
\put(-2,32){\makebox(0,0){1}}
\put(8,32){\makebox(0,0){2}} 
\put(18,32){\makebox(0,0){3}} 
\put(28,32){\makebox(0,0){4}} 
\put(38,32){\makebox(0,0){1}}
\put(20,-8){\makebox(0,0){(a)}}

%cyclic 334 lattice with L_y=3
\multiput(60,0)(10,0){5}{\circle*{2}}
\multiput(60,10)(10,0){5}{\circle*{2}}
\multiput(60,20)(10,0){5}{\circle*{2}}
\multiput(60,0)(10,0){5}{\line(0,1){20}}
\multiput(60,0)(0,10){3}{\line(1,0){40}}
\multiput(60,10)(10,0){4}{\line(1,1){10}}
\put(58,-2){\makebox(0,0){9}}
\put(68,-2){\makebox(0,0){10}}
\put(78,-2){\makebox(0,0){11}}
\put(88,-2){\makebox(0,0){12}}
\put(98,-2){\makebox(0,0){9}}
\put(58,12){\makebox(0,0){5}}
\put(68,12){\makebox(0,0){6}}
\put(78,12){\makebox(0,0){7}}
\put(88,12){\makebox(0,0){8}}
\put(98,12){\makebox(0,0){5}}
\put(58,22){\makebox(0,0){1}}
\put(68,22){\makebox(0,0){2}} 
\put(78,22){\makebox(0,0){3}} 
\put(88,22){\makebox(0,0){4}} 
\put(98,22){\makebox(0,0){1}}
\put(80,-8){\makebox(0,0){(b)}}
\end{picture}
\vspace*{3cm}

%cyclic 343 lattice with L_y=3
\begin{picture}(100,20)
\multiput(0,0)(10,0){5}{\circle*{2}}
\multiput(0,10)(10,0){5}{\circle*{2}}
\multiput(0,20)(10,0){5}{\circle*{2}}
\multiput(0,0)(10,0){5}{\line(0,1){20}}
\multiput(0,0)(0,20){2}{\line(1,0){40}}
\multiput(0,0)(10,0){4}{\line(1,1){10}}
\multiput(0,10)(10,0){4}{\line(1,1){10}}
\put(-2,-2){\makebox(0,0){9}}
\put(8,-2){\makebox(0,0){10}}
\put(18,-2){\makebox(0,0){11}}
\put(28,-2){\makebox(0,0){12}}
\put(38,-2){\makebox(0,0){9}}
\put(-2,12){\makebox(0,0){5}}
\put(8,12){\makebox(0,0){6}}
\put(18,12){\makebox(0,0){7}}
\put(28,12){\makebox(0,0){8}}
\put(38,12){\makebox(0,0){5}}
\put(-2,22){\makebox(0,0){1}}
\put(8,22){\makebox(0,0){2}} 
\put(18,22){\makebox(0,0){3}} 
\put(28,22){\makebox(0,0){4}} 
\put(38,22){\makebox(0,0){1}}
\put(20,-8){\makebox(0,0){(c)}}

%cyclic 35 lattice with L_y=3
\multiput(60,0)(10,0){5}{\circle*{2}}
\multiput(60,10)(10,0){5}{\circle*{2}}
\multiput(60,20)(10,0){5}{\circle*{2}}
\multiput(60,0)(10,0){5}{\line(0,1){20}}
\multiput(60,0)(0,20){2}{\line(1,0){40}}
\multiput(60,10)(10,0){4}{\line(1,1){10}}
\put(58,-2){\makebox(0,0){9}}
\put(68,-2){\makebox(0,0){10}}
\put(78,-2){\makebox(0,0){11}}
\put(88,-2){\makebox(0,0){12}}
\put(98,-2){\makebox(0,0){9}}
\put(58,12){\makebox(0,0){5}}
\put(68,12){\makebox(0,0){6}}
\put(78,12){\makebox(0,0){7}}
\put(88,12){\makebox(0,0){8}}
\put(98,12){\makebox(0,0){5}}
\put(58,22){\makebox(0,0){1}}
\put(68,22){\makebox(0,0){2}} 
\put(78,22){\makebox(0,0){3}} 
\put(88,22){\makebox(0,0){4}} 
\put(98,22){\makebox(0,0){1}}
\put(80,-8){\makebox(0,0){(d)}}
\end{picture}
\vspace*{3cm}

%cyclic 6 lattice with L_y=3
\begin{picture}(100,20)
\multiput(0,0)(10,0){5}{\circle*{2}}
\multiput(0,10)(10,0){5}{\circle*{2}}
\multiput(0,20)(10,0){5}{\circle*{2}}
\multiput(0,0)(10,0){5}{\line(0,1){20}}
\multiput(0,0)(0,20){2}{\line(1,0){40}}
\put(-2,-2){\makebox(0,0){9}}
\put(8,-2){\makebox(0,0){10}}
\put(18,-2){\makebox(0,0){11}}
\put(28,-2){\makebox(0,0){12}}
\put(38,-2){\makebox(0,0){9}}
\put(-2,12){\makebox(0,0){5}}
\put(8,12){\makebox(0,0){6}}
\put(18,12){\makebox(0,0){7}}
\put(28,12){\makebox(0,0){8}}
\put(38,12){\makebox(0,0){5}}
\put(-2,22){\makebox(0,0){1}}
\put(8,22){\makebox(0,0){2}} 
\put(18,22){\makebox(0,0){3}} 
\put(28,22){\makebox(0,0){4}} 
\put(38,22){\makebox(0,0){1}}
\put(20,-8){\makebox(0,0){(e)}}

%cyclic 44 lattice with L_y=3
\multiput(60,0)(10,0){5}{\circle*{2}}
\multiput(60,10)(10,0){5}{\circle*{2}}
\multiput(60,20)(10,0){5}{\circle*{2}}
\multiput(60,0)(10,0){5}{\line(0,1){20}}
\multiput(60,0)(0,20){2}{\line(1,0){40}}
\multiput(60,0)(10,0){4}{\line(1,2){10}}
\put(58,-2){\makebox(0,0){9}}
\put(68,-2){\makebox(0,0){10}}
\put(78,-2){\makebox(0,0){11}}
\put(88,-2){\makebox(0,0){12}}
\put(98,-2){\makebox(0,0){9}}
\put(58,12){\makebox(0,0){5}}
\put(68,12){\makebox(0,0){6}}
\put(78,12){\makebox(0,0){7}}
\put(88,12){\makebox(0,0){8}}
\put(98,12){\makebox(0,0){5}}
\put(58,22){\makebox(0,0){1}}
\put(68,22){\makebox(0,0){2}} 
\put(78,22){\makebox(0,0){3}} 
\put(88,22){\makebox(0,0){4}} 
\put(98,22){\makebox(0,0){1}}
\put(80,-8){\makebox(0,0){(f)}}
\end{picture}
\vspace*{1cm}

\begin{figure}[hbtp] 

\caption{\footnotesize{Illustrative strip graphs of the following lattices
with length $L_x=m=4$ and $(FBC_y,PBC_x)$ boundary condition (cyclic): (a)
$O_m$ lattice with $L_y=4$ (b) $Q_m$ lattice with $L_y=3$ (c) $S_m$
lattice with $L_y=3$ (d) $V_m$ lattice with $L_y=3$ (e) $X_{k,m}$ lattice
with $k=3$ (f) $Y_{k,m}$ lattice with $k=3$.}}

\label{illustration} 
\end{figure}

  In the present work we report exact solutions for chromatic polynomials
$P(G,q)$ for strips of lattices with arbitrarily great length $L_x$
vertices and finite width $L_y$. Consider $L_x$ sets of disjoint tree
graphs, $T_{L_y}$ with vertex set $v(T_{L_y}) = \{1,2,...,L_y\}$.  Denote
the edges joining each adjacent pair of $T_{L_y}$ and $T^\prime_{L_y}$ graphs 
as $e=\{v_iv^\prime_j|i,j=1,...,L_y\}$, where $v_i,v^\prime_j$ are vertices
of $T_{L_y}$ and $T^\prime_{L_y}$. Impose periodic boundary conditions in
the longitudinal direction, denoted $x$. The lattices in which we are 
interested here are (a) $L_y=4$ and $e=\{11,32,44\}$; (b) $L_y=3$ and
$e=\{11,21,22,33\}$; (c)  $L_y=3$ and $e=\{11,21,32,33\}$; (d) $L_y=3$ and
$e=\{11,21,33\}$; (e) $L_y=k$ and $e=\{11,kk\}$; (f) $L_y=k$ and
$e=\{11,k1,kk\}$. These families of graphs will be denoted as $O_m, Q_m,
S_m, V_m, X_{k,m}, Y_{k,m}$ in the present work, where $m=L_x$. 
Illustrative examples for these lattices with $m=4$ are displayed in Fig. 
\ref{illustration}.  The total number of vertices is $L_xL_y$ for all
these strips. We have given the chromatic
polynomial for the $L_y=3$ triangular lattice with $e=\{11,21,22,32,33\}$
and cyclic boundary condition before \cite{t}, and $Q_m$, $S_m$, and $V_m$
here are the corresponding lattice strips without one of the diagonal edges,
without the middle horizontal edge, and without both of these edges in
each subgraph, respectively. $X_{k,m}$ and $Y_{k,m}$ come from the
homeomorphic expansions on the vertical edges of the cyclic $L_y=2$ strips
of the square and triangular lattices.

A generic form for chromatic polynomials for recursively defined families
of graphs, of which strip graphs $G_s$ are special cases, is \cite{bkw}

\beq 
P((G_s)_m,q) = \sum_{j=1}^{N_{G_s,\lambda}}
c_{G_s,j}(q)(\lambda_{G_s,j}(q))^m \ ,
\label{pgsum} 
\eeq 
where $c_{G_s,j}(q)$ and the $N_{G_s,\lambda}$ terms $\lambda_{G_s,j}(q)$
depend on the type of strip graph $G_s$ but are independent of $m$.

The coefficients $c_{G_s,j}(q)$ that enter into the expressions for the
chromatic polynomial (\ref{pgsum}) for the cyclic strip of the families of
lattice considered here are

\beq
c^{(d)} = U_{2d}\Bigl ( \frac{\sqrt{q}}{2} \Bigr ) \ ,
\label{cd}   
\eeq
where $U_n(x)$ is the Chebyshev polynomial of the second kind, defined by

\beq
U_n(x) = \sum_{j=0}^{[\frac{n}{2}]} (-1)^j {n-j \choose j} (2x)^{n-2j}
\label{undef}
\eeq
as we found for the cyclic strips of square and triangular lattices
\cite{cf}.  Here in eq. (\ref{undef}) the notation $[\frac{n}{2}]$ in the
upper limit on the summand means the integral part of $\frac{n}{2}$. The
first few of these coefficients are

\beq
c^{(0)}=1 \ , \quad c^{(1)}=q-1 \ ,
\label{cd01}
\eeq

\beq
c^{(2)}=q^2-3q+1 \ ,
\label{cd2}  
\eeq
and

\beq
c^{(3)}=q^3-5q^2+6q-1 \ .
\label{cd3}
\eeq

When the coefficient $c^{(0)}$ appears in chromatic polynomials, we shall
simply write it as unity. 

For a given type of strip graph $G_s$, we denote the sum of the
coefficients $c_{G_s,j}$ as

\beq
C(G_s)=\sum_{j=1}^{N_{G_s,\lambda}} c_{G_s,j} \ .
\label{cgsum}
\eeq

It will also be convenient to define the following polynomial: 

\beq
D_k(q) = \frac{P(C_k,q)}{q(q-1)} = 
\sum_{s=0}^{k-2}(-1)^s {{k-1}\choose {s}} q^{k-2-s} \ ,
\label{dk}
\eeq
where $P(C_n,q)=(q-1)^n+(q-1)(-1)^n$ is the chromatic polynomial for the
circuit (cyclic) graph $C_n$ with $n$ vertices.  Some works on calculations of
chromatic polynomials for recursive families of graphs include 
\cite{bds}-\cite{tor4}.

\section{Family of $O_m$  Strips with $(FBC_y,PBC_x)$}

    In this section we give our solutions for the chromatic polynomials of the
$4 \times m $ strips of the $O_m$ lattice strip with cyclic boundary
conditions.  The chromatic number for this family is $\chi(O) = 3$.
We calculate the chromatic polynomials by iterated use of the
deletion-contraction theorem, via a generating function approach
\cite{strip,hs}, and a coloring matrix method \cite{b}.  We find
$N_{O,\lambda}=14$ and
\beq
P(O_m,FBC_y,PBC_x,q) = \sum_{j=1}^{14} c_{O,j} (\lambda_{O,j})^m \ ,
\label{pA}
\eeq
where $\lambda_{O,j}$'s for $1 \le j \le 13$ are roots of a cubic
equation, a sixth-degree equation, and a quartic equation.  Specifically, the
$\lambda_{O,j}$'s for $1 \le j \le 3$ are roots of the equation 

\beqs
& & \xi^3-(q^4-6q^3+16q^2-24q+17)\xi^2 \cr\cr
& & +(q-1)(q-2)(q^4-7q^3+20q^2-28q+18)\xi -(q-1)^2(q-2)^2=0
\label{eqA1}
\eeqs
while the $\lambda_{O,j}$'s for $4 \le j \le 9$ are roots of the equation 

\beqs
& & \xi^6+2(q^3-6q^2+14q-13)\xi^5+(q^2-5q+7)(q^4-9q^3+30q^2-47q+29)\xi^4
\cr\cr 
& & -(2q^8-30q^7+202q^6-796q^5+2001q^4-3278q^3+3420q^2-2084q+567)\xi^3
\cr\cr
& & +(q-1)(q^9-17q^8+130q^7-586q^6+1717q^5-3401q^4+4578q^3-4072q^2
      +2199q-557)\xi^2 \cr\cr
& & -2(q-1)^2(q-2)(q^5-9q^4+33q^3-61q^2+57q-23)\xi \cr\cr
& & +(q-1)^4(q-2)^2=0
\label{eqA2}
\eeqs
the $\lambda_{O,j}$'s for $10 \le j \le 13$ are roots of the equation

\beqs
& & \xi^4-(q^2-6q+10)\xi^3-(2q^3-13q^2+30q-23)\xi^2 \cr\cr
& & -(q-1)(q-3)(q^2-4q+5)\xi+(q-1)^2=0
\label{eqA3}
\eeqs
and finally, $\lambda_{O,14}=1$. 

    The corresponding coefficients are 
\beq
c_{O,j}=1 \quad {\rm for} \ \ 1 \le j \le 3
\label{cA13}
\eeq

\beq  
c_{O,j}=c^{(1)} \quad {\rm for} \ \ 4 \le j \le 9
\label{cA49}
\eeq

\beq  
c_{O,j}=c^{(2)} \quad {\rm for} \ \ 10 \le j \le 13
\label{cA1013}
\eeq
and 
\beq  
c_{O,14}=c^{(3)} \ .
\label{cA14}
\eeq
Here the coefficients have the form of the Chebyshev polynomial of the
second kind given in (\ref{undef})-(\ref{cd3}).  There is also a zero
eigenvalue $\lambda_{O,15}=0$ with coefficient
$c_{O,15}=q(q-1)(q^2-3q+1)$, which does not contribute to the chromatic
polynomial (\ref{pA}). The sum of all of the coefficients is equal to
$P(T_4,q)=q(q-1)^3$ if we include $c_{O,15}$, and is equal to $q^2(q-1)$
if $c_{O,15}$ is not included. 

\begin{figure}[hbtp]
\centering
\leavevmode
\epsfxsize=4.0in
\begin{center}
\leavevmode
\epsffile{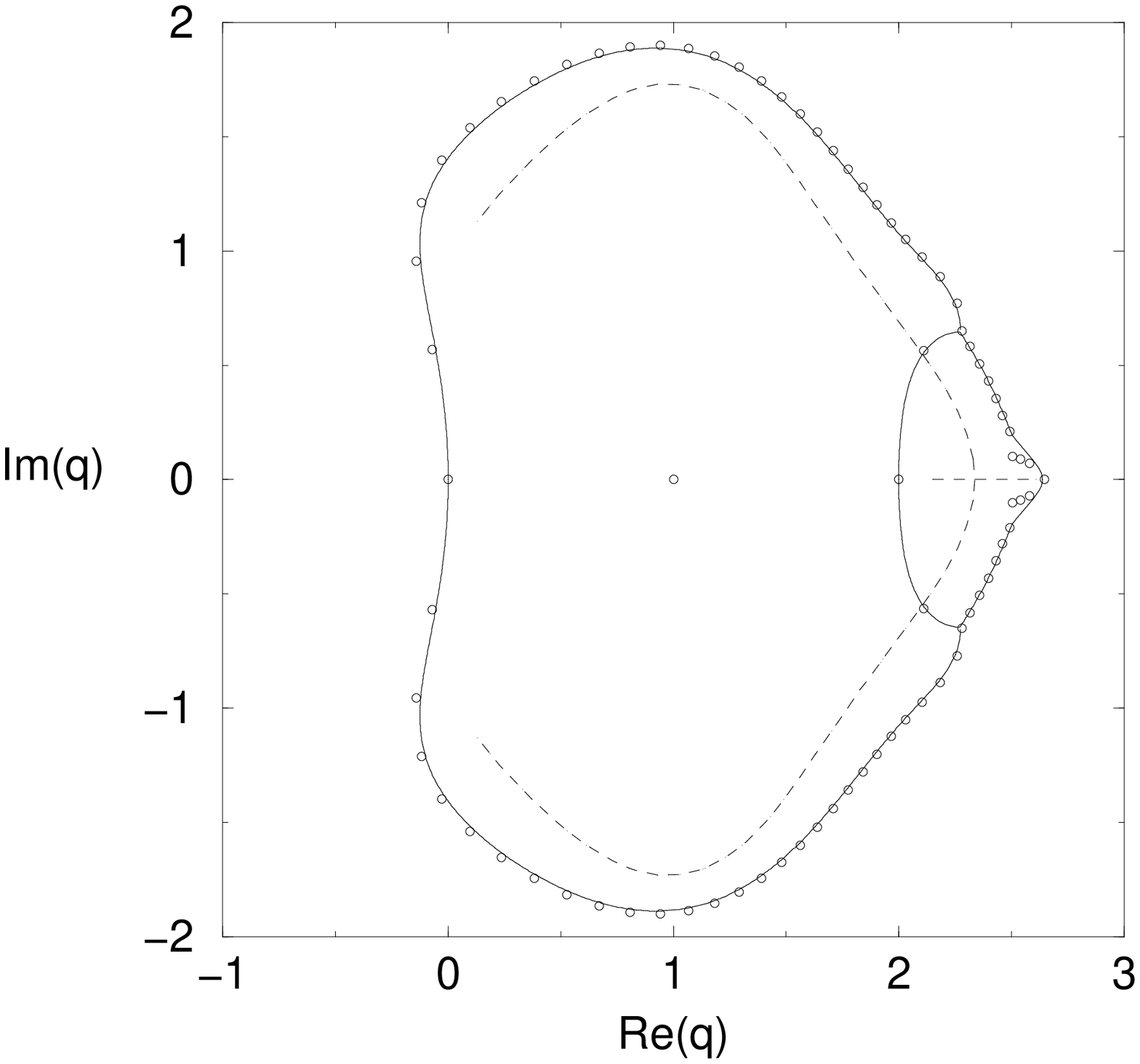}
\end{center}
\caption{\footnotesize{The locus ${\cal B}$ for the $m \to \infty$ limit of the
$O_m$ family of cyclic strip graphs (solid curve).  For comparison, we show 
chromatic zeros for the cyclic $O_m$ strip graph with $m=20$, i.e., $n=80$ 
vertices. The dashed line is the locus of solutions of the degeneracy equation
involving only the roots of the cubic eq. (\ref{eqA1}).}}
\label{Azeros}
\end{figure}

The locus ${\cal B}$ separates the $q$ plane into three main regions.  The
outermost one, region $R_1$, extends to infinite $|q|$ and includes the
intervals $q \ge q_c$ and $q \le 0$ on the real $q$ axis.  Here, 
\beq
q_c(O,FBC_y,PBC_x) \simeq 2.638342 
\label{qcA}
\eeq
Region $R_2$
includes the real interval $2 \le q \le q_c$, while region $R_3$ includes
the real interval $0 \le q \le 2$.  In regions $R_i$, $1 \le i \le 3$, the
dominant terms are, respectively, the root with maximal magnitude of (1)
the cubic equation (\ref{eqA1}), (2) the quartic equation (\ref{eqA3}),
(3)  the sixth-degree equation (\ref{eqA2}). Our previous calculations for
various families of graphs \cite{wcy,t} have shown that ${\cal B}$ can
include pairs of extremely small complex-conjugate regions. We have not
made an exhaustive search for these in the present case. 
The locus ${\cal B}$ for this family (defined in the limit $n \to \infty$) 
and, for comparison, chromatic zeros for a typical long finite $O_m$ strip,
with $L_x=m=20$, are shown in Fig. \ref{Azeros}.  With this value of
$m$, the chromatic zeros lie close to the boundary ${\cal B}$.  The locus
${\cal B}$ crosses the real $q$-axis at $q=0,2$ and $q_c$.

Earlier, it was shown that the locus ${\cal B}$ can have support for $Re(q) \le
0$ in \cite{wcy} for square lattice strips and in \cite{t} for triangular
lattice with cyclic or M\"obius boundary conditions.  These families have
respective girths equal to 4 and 3, respectively, for $L_x$ large enough to
avoid degenerate cases.  Chromatic zeros with negative real parts were also
found for several other families of graphs \cite{read91,hs}, including those
that yielded chromatic zeros with arbitrarily large magnitude $|q|$
\cite{wa2,sokal}. Here we find that ${\cal B}(O)$ also has support for $Re(q)
\le 0$ for the present $m \to \infty$ limit of) the $O_m$ family, which has
girth 5. Although the chromatic roots of regular dodecahedron, denoted as $O_5$
here, does not have any negative real part, where the smallest one is about
0.0082126, except for the trivial root $q=0$, the chromatic roots of $O_m$ for
$m \ge 6$ do include some with negative real parts. The largest magnitude of
the negative real part increases as $m$ increases, and approaches the limit of
${\cal B}$ up to $q \simeq -0.125$. The density of zeros is observed to be
relatively small on the parts of ${\cal B}$ extending through $q=0$ and 2,
separating $R_3$ from $R_1$ and $R_2$ near the real $q$-axis.

Partial results for the chromatic polynomial of the $O_m$ family were reported
in \cite{lse0007}, where this family was denoted as $D_n$.  Where they overlap,
some of the results in our complete calculation differ with those in 
\cite{lse0007} \cite{brnote}.  

\section{Family of $Q_m$  Strips with $(FBC_y,PBC_x)$}

    In this section we consider the chromatic polynomials of the $3 \times m$
strips of the $Q_m$ lattice with $(FBC_y,PBC_x)$, i.e. cyclic, boundary
conditions. This family of graphs has one more diagonal edge in each subgraph
than the $L_y=3$ square lattice with cyclic boundary condition \cite{wcyl}, and
one less diagonal edge than the $L_y=3$ triangular lattice with the same
boundary condition \cite{t}.  We have $N_{Q,\lambda}=10$, and

\beq
P(Q_m,FBC_y,PBC_x,q) = \sum_{j=1}^{10} c_{Q,j} (\lambda_{Q,j})^m \ ,
\label{pB}
\eeq
where

\beq
\lambda_{Q,(1,2)}=\frac{1}{2}\Bigl ( q^3 - 6q^2 + 14q - 13 \pm
         \sqrt{(q-3)(q^2-5q+7)(q^3-4q^2+6q-5)} \ \Bigr ) \ .
\label{lamB12}
\eeq

The $\lambda_{Q,j}$ for $3 \le j \le 6$ are the roots of the quartic equation

\beqs
& & \xi^4 + (3q^2-15q+20)\xi^3+(3q^4-30q^3+114q^2-191q+117)\xi^2 \cr\cr
& & +(q-2)(q^5-14q^4+76q^3-199q^2+251q-121)\xi - (q-2)^5(q^2-5q+5) = 0
\label{lamB36}
\eeqs

\beq
\lambda_{Q,(7,8)}= q-3 \pm \sqrt{4-q}
\label{lamB78}
\eeq
and $\lambda_{Q,9} = q-2$, $\lambda_{Q,10} = -1$. The coefficients are

\beq
c_{Q,j}=1 \quad {\rm for} \ \ 1 \le j \le 2
\label{cB12}
\eeq 

\beq
c_{Q,j}=c^{(1)} \quad {\rm for} \ \ 3 \le j \le 6
\label{cB36}
\eeq

\beq
c_{Q,j}=c^{(2)} \quad {\rm for} \ \ 7 \le j \le 9
\label{cB79}
\eeq
and 

\beq
c_{Q,10}=c^{(3)} \ .
\label{cB10}
\eeq

Again only one type of polynomial of each degree in $q$ occurs in the
coefficients, and the number of $\lambda$'s with coefficients $c^{(d)}$,
which is denoted as $n_P(L_y,d)$ in \cite{cf}, are exactly the same as
those for the square and triangular lattices with $L_y=3$:  $n_P(3,0)=2,
n_P(3,1)=4, n_P(3,2)=3, n_P(3,3)=1$. Therefore, the sum of the
coefficients is

\beq
C(Q, FBC_y, PBC_x) = \sum_{j=1}^{10} c_{Q,j} = q(q-1)^2 \ .
\label{cbsum}
\eeq

For the family $Q_m$ of lattice strips, we find that $q_c=3$; at this point all
the $\lambda_Q$'s have the same magnitude 1. The locus ${\cal B}$ crosses the
real $q$-axis at $q=0,2$ and $q_c=3$, and separates the $q$ plane into five
main regions.  Region $R_1$ extends to infinite $|q|$ and includes the
intervals $q \ge 3$ and $q \le 0$ on the real $q$ axis.  Region $R_2$ includes
the real interval $2 \le q \le 3$, and region $R_3$ includes the real interval
$0 \le q \le 2$. There is also a pair of complex-conjugate regions $R_4,
R_4^*$ centered at approximately $q \simeq 2.55 \pm 1.30i$. We have not
made an exhaustive search for the tiny regions in this case.  The dominant
terms in regions $R_1$ and $R_2$ are $\lambda_{Q,1}$ in (\ref{lamB12}) and
$\lambda_{Q,7}$ in (\ref{lamB78}), while regions $R_3$ and $R_4, R_4^*$
are both dominated by the roots of the quartic equation (\ref{lamB36})
with maximal magnitude.  The locus ${\cal B}$ has support for $Re(q) \ge
0$, and the density of zeros is smaller on the parts of ${\cal B}$
extending through $q=0$ and 2, separating $R_3$ from $R_1$ and $R_2$. 

\begin{figure}[hbtp]
\centering
\leavevmode
\epsfxsize=4.0in
\begin{center}
\leavevmode
\epsffile{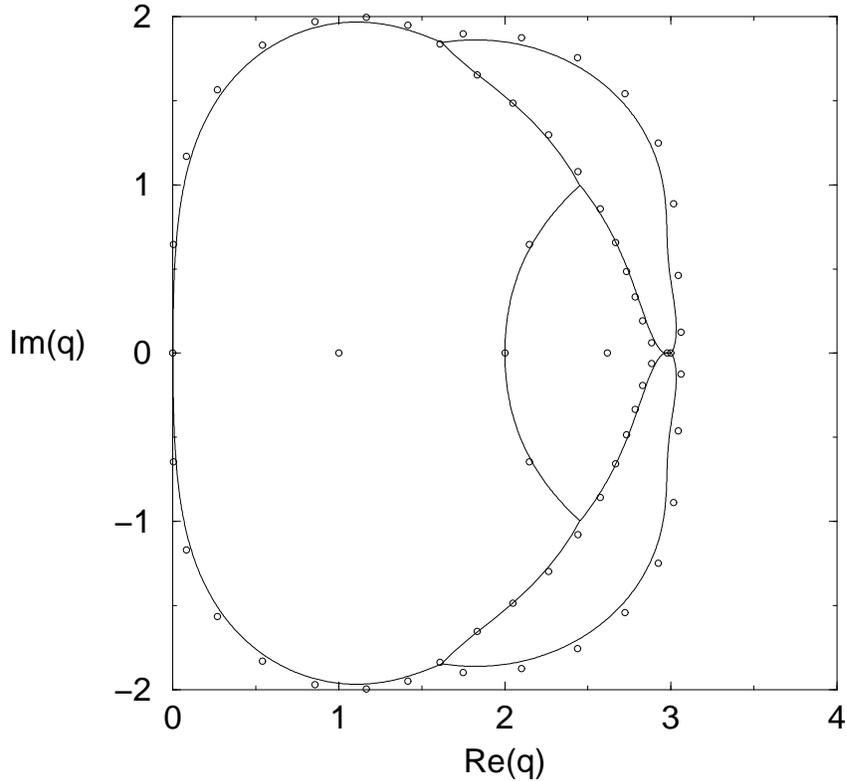}
\end{center}
\caption{\footnotesize{Locus ${\cal B}$ for the $m \to \infty$ limit of the
family of cyclic graphs $Q_m$.  For comparison, chromatic zeros for 
$Q_m$ with $m=20$ (i.e., $n=60$) are shown.}}
\label{Bzeros}
\end{figure}

The locus ${\cal B}$ and chromatic zeros for the $L_x=m=20$ cyclic graph
of the $Q_m$ lattice are shown in Fig. \ref{Bzeros}. The chromatic zeros
lie close to the boundary ${\cal B}$ and indicate its position.

\section{Family of $S_m$ Strips with $(FBC_y,PBC_x)$}

The $3 \times m$ strips of the $S_m$ lattice with $(FBC_y,PBC_x)=cyclic$
boundary conditions have one less middle horizontal edge in each subgraph than
the $L_y=3$ triangular lattice with the same boundary conditions \cite{t}. For
the solutions of the chromatic polynomial, we find $N_{S,\lambda}=6$ and

\beq
P(S_m,FBC_y,PBC_x,q) = \sum_{j=1}^{6} c_{S,j} (\lambda_{S,j})^m \ ,
\label{pC}
\eeq
where

\beq
\lambda_{S,(1,2)}=\frac{q-2}{2}\Bigl ( q^2 - 4q + 6 \pm
         \sqrt{q^4-8q^3+24q^2-36q+28} \ \Bigr ) \ ,
\label{lamC12}
\eeq

\beq
\lambda_{S,(3,4)}=\frac{-1}{2}\Bigl ( (q-3)(q-4) \pm
         \sqrt{q^4-10q^3+49q^2-120q+112} \ \Bigr ) \ ,
\label{lamC34}
\eeq
and $\lambda_{S,5} = -(q-2)^2$, $\lambda_{S,6} = q-4$. The coefficients
are

\beq
c_{S,(1,2)}=1
\label{cC12}
\eeq 

\beq
c_{S,j}=c^{(1)} \quad {\rm for} \ \ 3 \le j \le 5
\label{cC35}
\eeq
and

\beq
c_{S,6}=c^{(2)} \ .
\label{cC6}
\eeq
These coefficients have the form of Chebyshev polynomials of the second kind,
as given above.  There is also a zero eigenvalue $\lambda_{S,7}=0$
with coefficient $c_{S,7}=q(q^2-3q+1)$ which does not contribute to the
chromatic polynomial (\ref{pC}). The sum of the coefficients is equal to
$P(T_3,q)=q(q-1)^2$ if we include $c_{S,7}$, and is equal to $q^2$ if $c_{S,7}$
is not included.

\begin{figure}[hbtp]
\centering
\leavevmode
\epsfxsize=4.0in
\begin{center}
\leavevmode
\epsffile{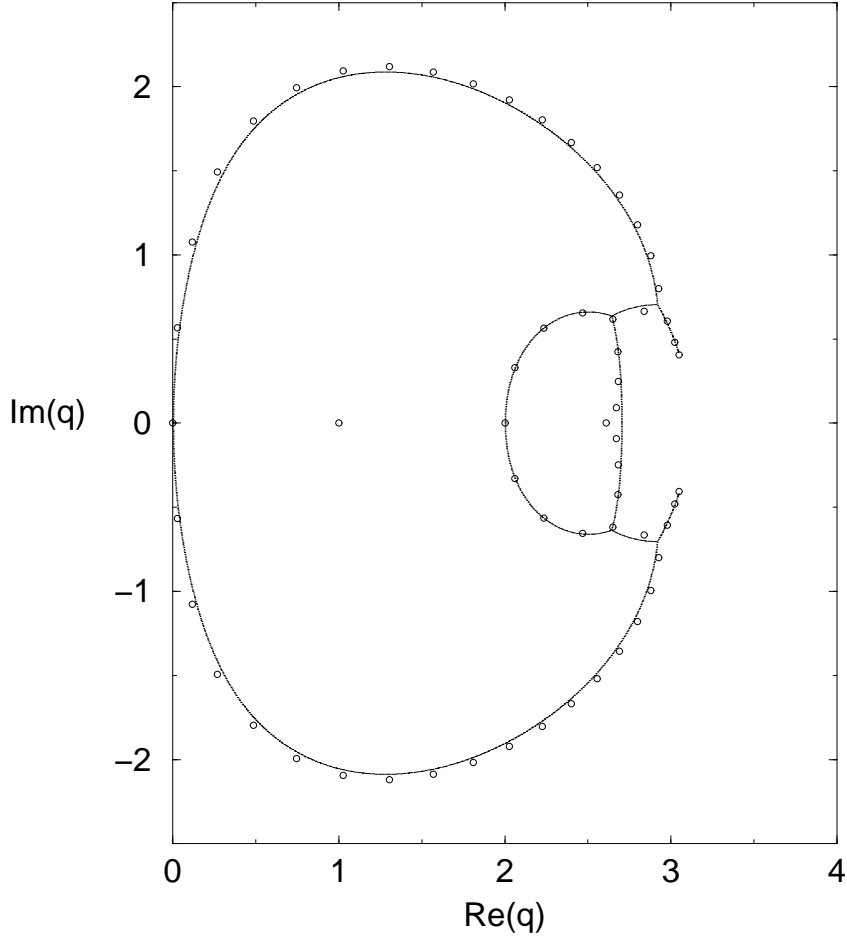}
\end{center}
\caption{\footnotesize{Locus ${\cal B}$ for the $m \to \infty$ limit of the
cyclic family $S_m$.  For comparison, chromatic zeros for the graph $S_m$ with
$m=20$ (i.e., $n=60$) are shown.}}
\label{Czeros}
\end{figure}

The locus ${\cal B}$ crosses the real $q$-axis at $q=0,2$ and $q_c$, where
\beq
q_c(S, FBC_y, PBC_x) = 2.702873 \ ,
\label{qcC}
\eeq
which is the real solution of equation $2q^3-11q^2+24q-24=0$.  
This locus separates the $q$ plane into three regions.  Region $R_1$ extends to
infinite $|q|$ and includes the intervals $q \ge q_c$ and $q \le 0$ on the
real $q$ axis.  Region $R_2$ includes the real interval $2 \le q \le q_c$,
while region $R_3$ includes the real interval $0 \le q \le 2$. The
dominant terms in regions $R_i$, $1\le i \le 3$ are (i) $\lambda_{S,1}$
in (\ref{lamC12}), (ii) $\lambda_{S,6}$, and (iii) $\lambda_{S,3}$ in
(\ref{lamC34}), respectively. Therefore, $q_c$ for the $m \to \infty$ limit of
this lattice strip $S_m$ is given by the degeneracy between $|\lambda_{S,1}|$ 
and $|\lambda_{S,6}|$, which yields the value in (\ref{qcC}). 
The locus ${\cal B}$ and the chromatic zeros for the $L_x=m=20$ cyclic
graph of the $S_m$ lattice strip are shown in Fig. \ref{Czeros}.  This locus
has support
for $Re(q) \ge 0$. The locus ${\cal B}(S)$ not only has closed regions
$R_1$ to $R_3$, but also has a pair of arcs extending from $q\simeq 2.9
\pm 0.7i$ at the boundary between $R_1$ and $R_3$ to $q\simeq 3.05
\pm 0.396i$. 

\section{Family of $V_m$  Strips with $(FBC_y,PBC_x)$}

We next proceed to the $3 \times m$ strips of the $V_m$ lattice with
$(FBC_y,PBC_x)$, which have one less horizontal edge in each subgraph than the
$Q_m$ lattice, or equivalently one less diagonal edge in each subgraph than the
$S_m$ lattice. We find $N_{V,\lambda}=6$ as for lattice $S_m$, and

\beq
P(V_m,FBC_y,PBC_x,q) = \sum_{j=1}^{6} c_{V,j} (\lambda_{V,j})^m \ ,
\label{pD}
\eeq
where

\beq
\lambda_{V,(1,2)}=\frac{q-2}{2}\Bigl ( q^2 - 3q + 4 \pm
         \sqrt{(q-2)(q-3)(q^2-q+2)} \ \Bigr ) \ ,
\label{lamD12}
\eeq

$\lambda_{V,j}$ for $j=3 \le j \le 5$ are roots of the cubic equation 

\beqs
& & \xi^3+(2q^2-9q+11)\xi^2 \cr\cr
& & +(q-2)(q^3-8q^2+20q-14)\xi -(q-1)^2(q-2)^2(q-3)=0
\label{lamD35}
\eeqs
and $\lambda_{V,6} = q-3$. The corresponding coefficients are

\beq
c_{V,(1,2)}=1
\label{cD12}
\eeq 

\beq
c_{V,j}=c^{(1)} \quad {\rm for} \ \ 3 \le j \le 5
\label{cD35}
\eeq
and

\beq
c_{V,6}=c^{(2)} \ .
\label{cD6}
\eeq
These coefficients are the same as the coefficients for the family of lattice
strips $S_m$, and the zero eigenvalue $\lambda_{V,7}=0$ has the same
coefficient $c_{V,7}=q(q^2-3q+1)$. Thus, the sum of the coefficients is equal
to $P(T_3,q)=q(q-1)^2$ if we include $c_{V,7}$, and is $q^2$ if $c_{V,7}$ if
not included.

\begin{figure}[hbtp]
\centering
\leavevmode
\epsfxsize=4.0in
\begin{center}
\leavevmode
\epsffile{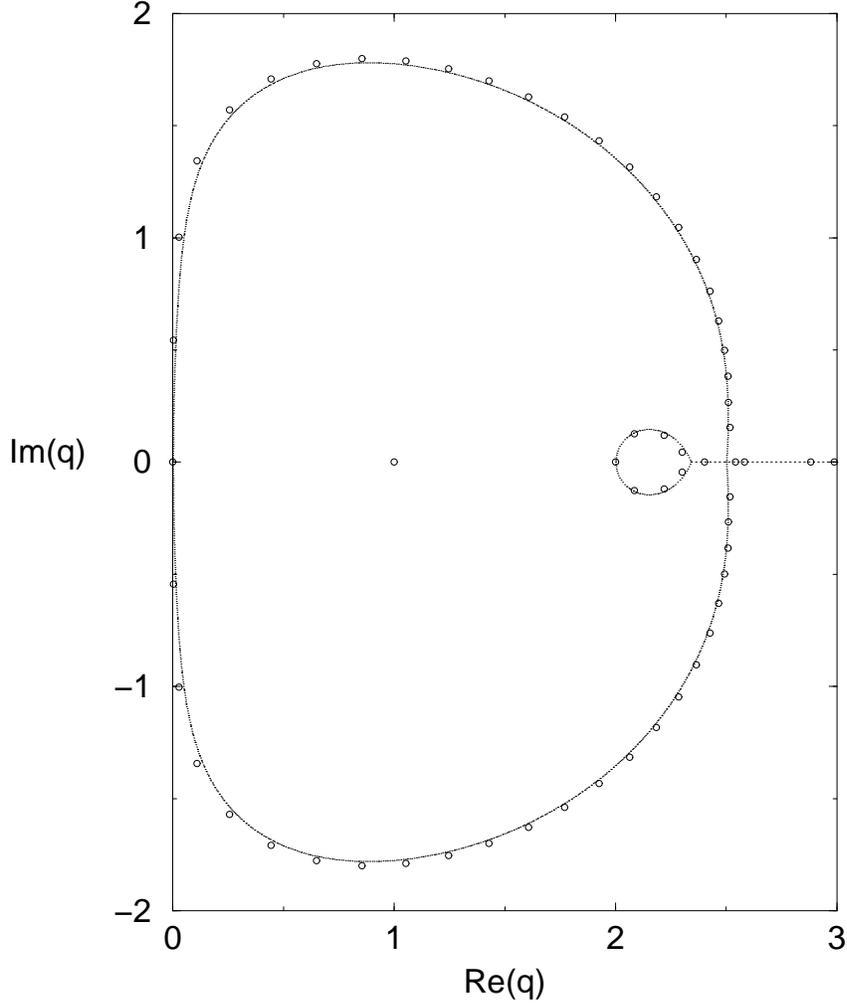}
\end{center}
\caption{\footnotesize{Locus ${\cal B}$ for the $m \to \infty$ limit of the
cyclic $V_m$ family.  For comparison, chromatic zeros for the cyclic $V_m$
graph with $m=20$ (i.e., $n=60$) are shown.}}
\label{Dzeros}
\end{figure}

The locus ${\cal B}$ has support for $Re(q) \ge 0$, and consists of (1) 
an outer closed region which crosses the real $q$-axis at $q=0,
\frac{5}{2}$; (2)  an inner closed region which crosses the real $q$-axis
at $q=2$ and $q\simeq 2.341$; and (3) a finite line segment on the the real
$q$-axis extending from $q\simeq 2.341$ to $q=3$.  Thus, the $q$ plane is
separated into three regions and $q_c=3$.  Region $R_1$ is the exterior part of
the outer closed region, and extends to infinite $|q|$. Region $R_3$ is the
interior part of the inner closed region, and region $R_2$ includes the
interior part of the outer closed region and exterior part of $R_3$.  The
dominant terms in regions $R_i$, $1\le i \le 3$ are (i) $\lambda_{V,1}$ in
(\ref{lamD12}), (ii) the roots of eq. (\ref{lamD35}) with largest magnitude,
and (iii) $\lambda_{V,6}$, respectively.  The locus ${\cal B}$ for the $m \to
\infty$ limit of the $V_m$ family and the
chromatic zeros for the $L_x=m=20$ $V_m$ graph are shown
in Fig. \ref{Dzeros}. As was true for the $S_m$ lattice, The locus ${\cal
B}(V)$ has both closed regions and finite line segment.

\section{Family of $X_{k,m}$  Strips with $(FBC_y,PBC_x)$}

The homeomorphic expansion of the $L_y=2$ square lattice with cyclic
boundary condition on the horizontal edges has been studied before
\cite{nec}. In this section we consider the homeomorphic expansion of the
$L_y=2$ square lattice with cyclic boundary condition on the vertical
edges, so that there are $k$ vertices on the transverse direction, i.e. we
make the insertion of $k-2$ degree-2 vertices on all of the vertical edges
of the strip. This is equivalent to the $L_y=k$ square lattice having the
same boundary condition without all the internal horizontal edges. We find
$N_{X,\lambda}=6$ as for lattice $V_m$, and

\beq
P(X_{k,m},FBC_y,PBC_x,q) = \sum_{j=1}^{6} c_{X,j} (\lambda_{X_k,j})^m \ ,
\label{pE}
\eeq
where $\lambda_{X_k,(1,2)}$ are the roots of the quadratic equation,

\beqs
& & \xi^2+\frac{(-1)^k}{q}\Bigl ( (q^2-2q+2)(1-q)^{(k-1)} + q-2 \Bigr )\xi
\cr\cr
& & +\frac{(q-1)^3}{q^2}\Bigl ( (q-1)^{(2k-3)} - (q-2)(1-q)^{(k-2)}-1
\Bigr ) =0  \ .
\label{lamE12}
\eeqs

The $\lambda_{X_k,(3,4)}$ are the roots of another quadratic equation,

\beqs
& & \xi^2+\frac{(-1)^k}{q}\Bigl ( (2-q)(1-q)^{(k-1)} - 2 \Bigr )\xi \cr\cr
& & -\frac{1}{q^2}\Bigl ( (q-1)^{(2k-1)} - (q-2)(1-q)^k - (q-1)^2 \Bigr )
=0 
\label{lamE34}
\eeqs

\beq
\lambda_{X_k,5} = -(q-1)D_k(q) \ ,
\label{lamE5}
\eeq
and 

\beq
\lambda_{X_k,6} = D_k(q) \ .
\label{lamE6}
\eeq
where $D_k(q)$ is defined in (\ref{dk}). In eqs. (\ref{lamE12}) and
(\ref{lamE34}), the coefficient functions are polynomials for any positive
integer $k$.

The corresponding coefficients are

\beq
c_{X,(1,2)}=1
\label{cE12}
\eeq 

\beq
c_{X,j}=c^{(1)} \quad {\rm for} \ \ 3 \le j \le 5
\label{cE35}
\eeq
and

\beq
c_{X,6}=c^{(2)} \ .
\label{cE6}
\eeq
These coefficients are the same as the coefficients for the lattices $S$
and $V$. Therefore, the sum of the coefficients is equal to $q^2$ if we
do not consider the zero eigenvalue. The chromatic number is 2 and 3 when
$L_x$ is even and odd, respectively, independent of $k$. 

This lattice is the only one we consider here which does not contain a seam
when we impose periodic longitudinal boundary conditions with an
orientation-reversal (M\"obius). We have $\lambda_{X_k,mb,j} =
\lambda_{X_k,j}, 1 \le j \le 6$, and the corresponding coefficients are 

\beq
c_{X,mb,(1,2)}=1
\label{cEmb12}
\eeq 

\beq
c_{X,mb,(3,4)}=c^{(1)}
\label{cEmb34}
\eeq

\beq
c_{X,mb,5}=-c^{(1)}
\label{cEmb5}
\eeq
and

\beq
c_{X,mb,6}=-1 \ .
\label{cEmb6}
\eeq
The changes of the coefficients when we change the longitudinal boundary
condition from cyclic to M\"obius follow the rules given in Theorem 6 of
\cite{cf}.  The chromatic number for the M\"obius strips of lattice
$X_{k,m}$ is 2 as $(k,L_x)=(o,e),(e,o)$, and 3 as $(k,L_x)=(o,o), (e,e)$,
where $e$ and $o$ denote even and odd values for $k$ and $L_x$. 

The fact that the $\lambda_j$'s for a M\"obius strip must be the same as
those for the cyclic strip of the same width and lattice type was proved
in \cite{pm}; this also proves, {\it a fortiori}, that (i) the total
number, $N_\lambda$, of $\lambda_j$'s, and (ii) the continuous nonanalytic
locus ${\cal B}$, including the point $q_c$, are the same for the cyclic
and M\"obius strips.

\begin{figure}[hbtp]
\centering
\leavevmode
\epsfxsize=4.0in
\begin{center}
\leavevmode
\epsffile{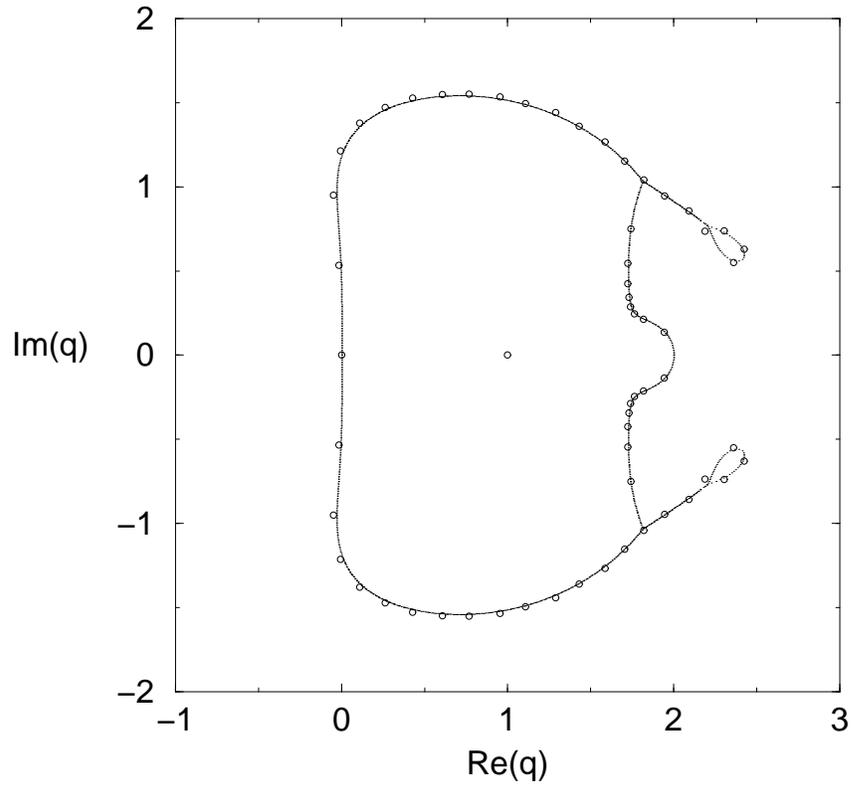}
\end{center}
\caption{\footnotesize{Locus ${\cal B}$ for the $m \to \infty$ limit of the
cyclic $X_{3,m}$ family and chromatic zeros for the cyclic $X_{3,m}$ graph 
with $m=20$ (i.e., $n=60$).}}
\label{E3zeros}
\end{figure}

\begin{figure}[hbtp]
\centering
\leavevmode
\epsfxsize=4.0in
\begin{center}
\leavevmode
\epsffile{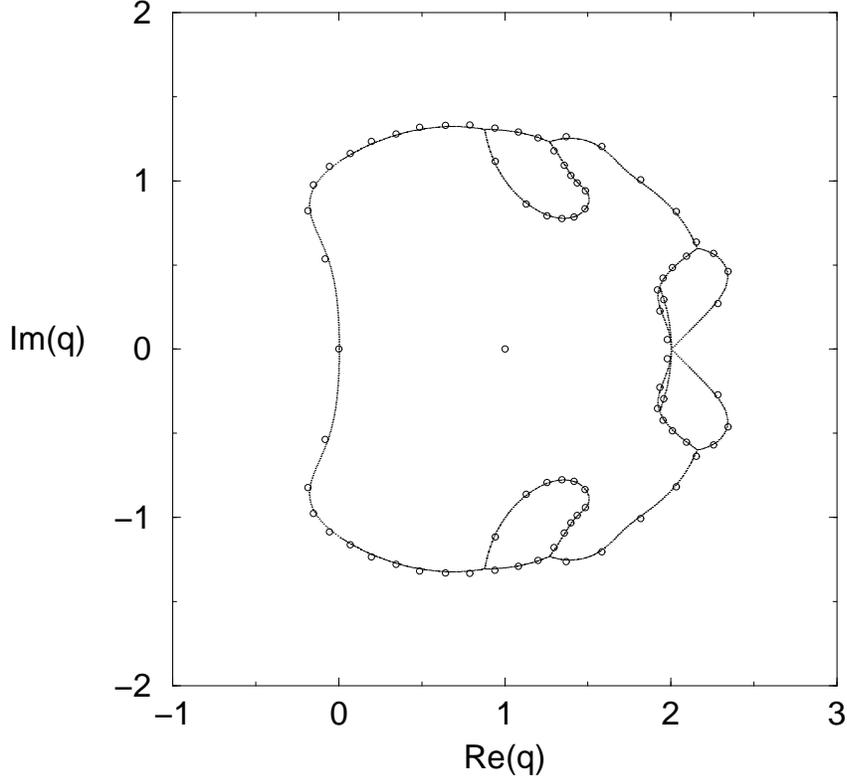}
\end{center}
\caption{\footnotesize{Locus ${\cal B}$ for the $m \to \infty$ limit of the
cyclic $X_{4,m}$ family and chromatic zeros for the cyclic $X_{4,m}$ graph
with $m=20$ (i.e., $n=80$).}}
\label{E4zeros}
\end{figure}

The locus ${\cal B}$ and the chromatic zeros for the $L_x=m=20$ cyclic
graph of the $X_{k,m}$ lattice with $k=3, 4$ are shown in Fig.
\ref{E3zeros} and \ref{E4zeros}. Except for $k=2$ which is the ladder
graph, we find the locus ${\cal B}(X_k)$ for $k \ge 3$ has support for
$Re(q) \le 0$, and crosses the real $q$-axis at $q=0$ and 2. The two main
regions are region $R_1$ includes the intervals $q \ge 2$ and $q \le 0$ on
the real $q$ axis, and region $R_2$ includes the real interval $0 \le q
\le 2$. The dominant terms in regions $R_1$ and $R_2$ are roots of
(\ref{lamE12}) and (\ref{lamE34}), respectively.

For odd $k$, there is also a pair of complex-conjugate regions $R_3,
R_3^*$ on the upper right and lower right of $q=q_c=2$, which are connected
to $R_2$ by arcs. The dominant term in these regions is
$\lambda_{X_{odd},3}$. The locus ${\cal B}(X_{odd})$ around $q_c$ has the
form of a circular arc which bends to the left before reaching a pair of
complex-conjugate roots of the discriminant of eq. (\ref{lamE34}). For
even $k$, a general feature is that there are two contiguous pairs of
complex-conjugate regions $R_3, R_3^*$ on the upper right and lower right
of $q=2$ and $R_4, R_4^*$ on the upper left and lower left of $q=2$, and
both of them touch the real $q$-axis at $q=2$. The dominant terms in $R_3,
R_3^*$ and $R_4, R_4^*$ are $\lambda_{X_{even},5}$ and
$\lambda_{X_{even},6}$, respectively.

\begin{figure}[hbtp]
\centering  
\leavevmode
\epsfxsize=4.0in
\begin{center}
\leavevmode
\epsffile{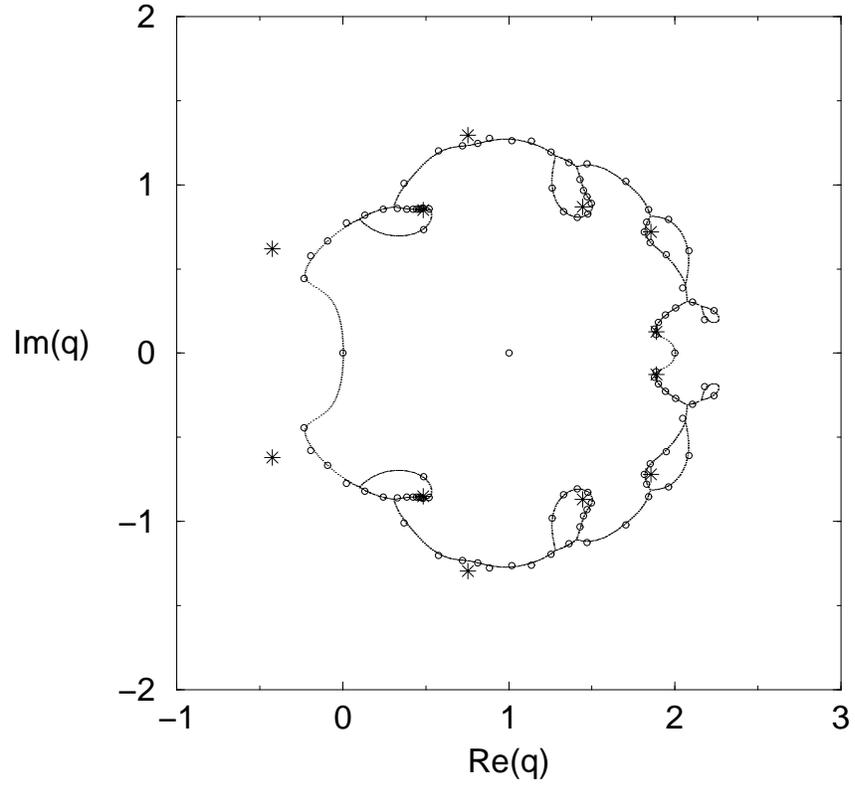}
\end{center}
\caption{\footnotesize{Locus ${\cal B}$ for the $m \to \infty$ limit of the
family of cyclic graphs $X_{7,m}$ and chromatic zeros for the cyclic
graph $X_{7,m}$ with $m=15$ (i.e., $n=105$). The roots of the discriminant of
eq. (\ref{lamE34}) are indicated with $*$. }}
\label{E7zeros}
\end{figure}

\begin{figure}[hbtp]
\centering  
\leavevmode
\epsfxsize=4.0in
\begin{center}
\leavevmode
\epsffile{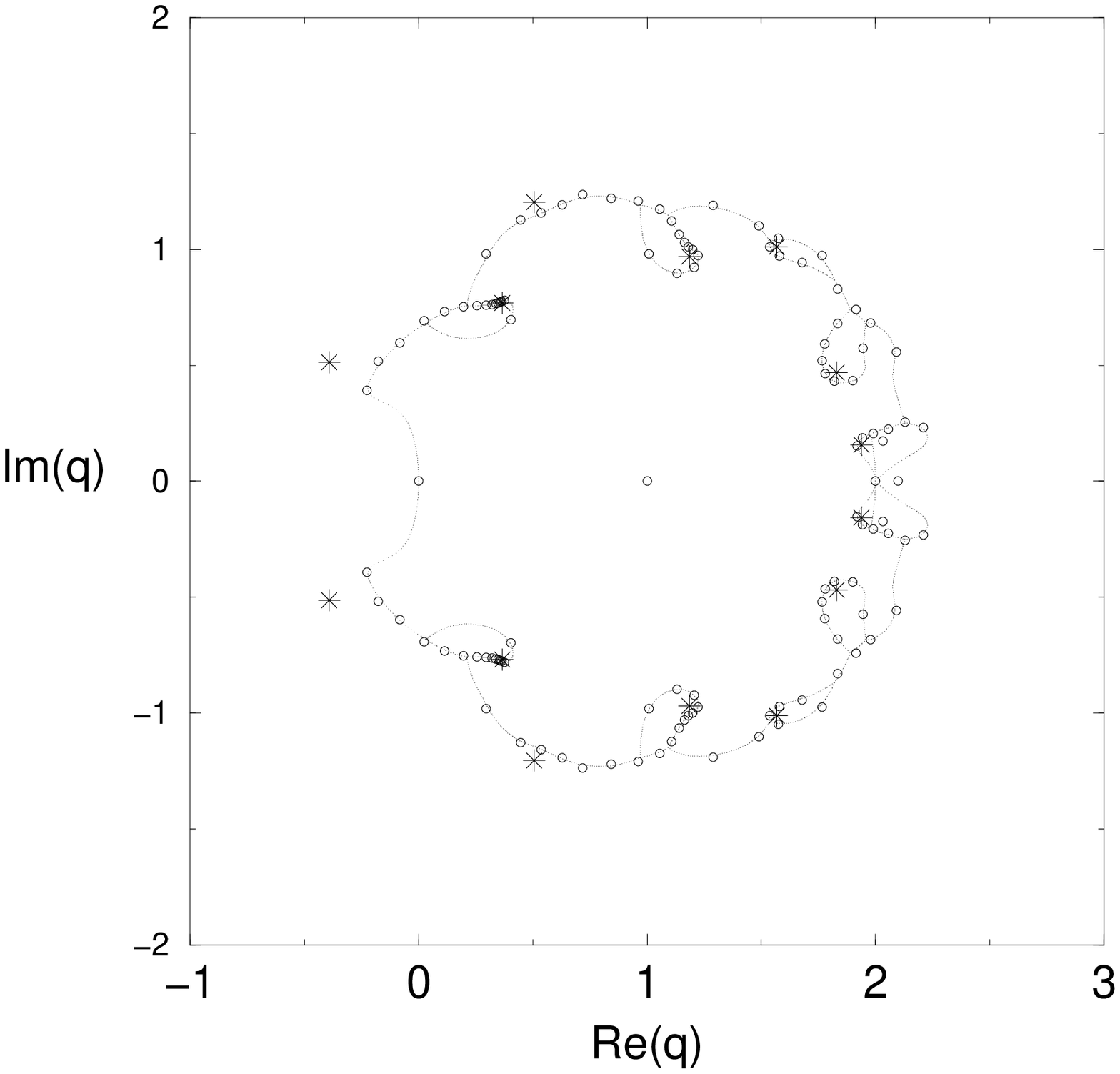}
\end{center}
\caption{\footnotesize{Locus ${\cal B}$ for the $m \to \infty$ limit of the
family of cyclic graphs $X_{8,m}$ and chromatic zeros for the cyclic
graph $X_{8,m}$ with $m=15$ (i.e., $n=120$). The roots of the discriminant of
eq. (\ref{lamE34}) are indicated with $*$.}}
\label{E8zeros}
\end{figure}

\begin{figure}[hbtp]
\centering  
\leavevmode
\epsfxsize=4.0in
\begin{center}
\leavevmode
\epsffile{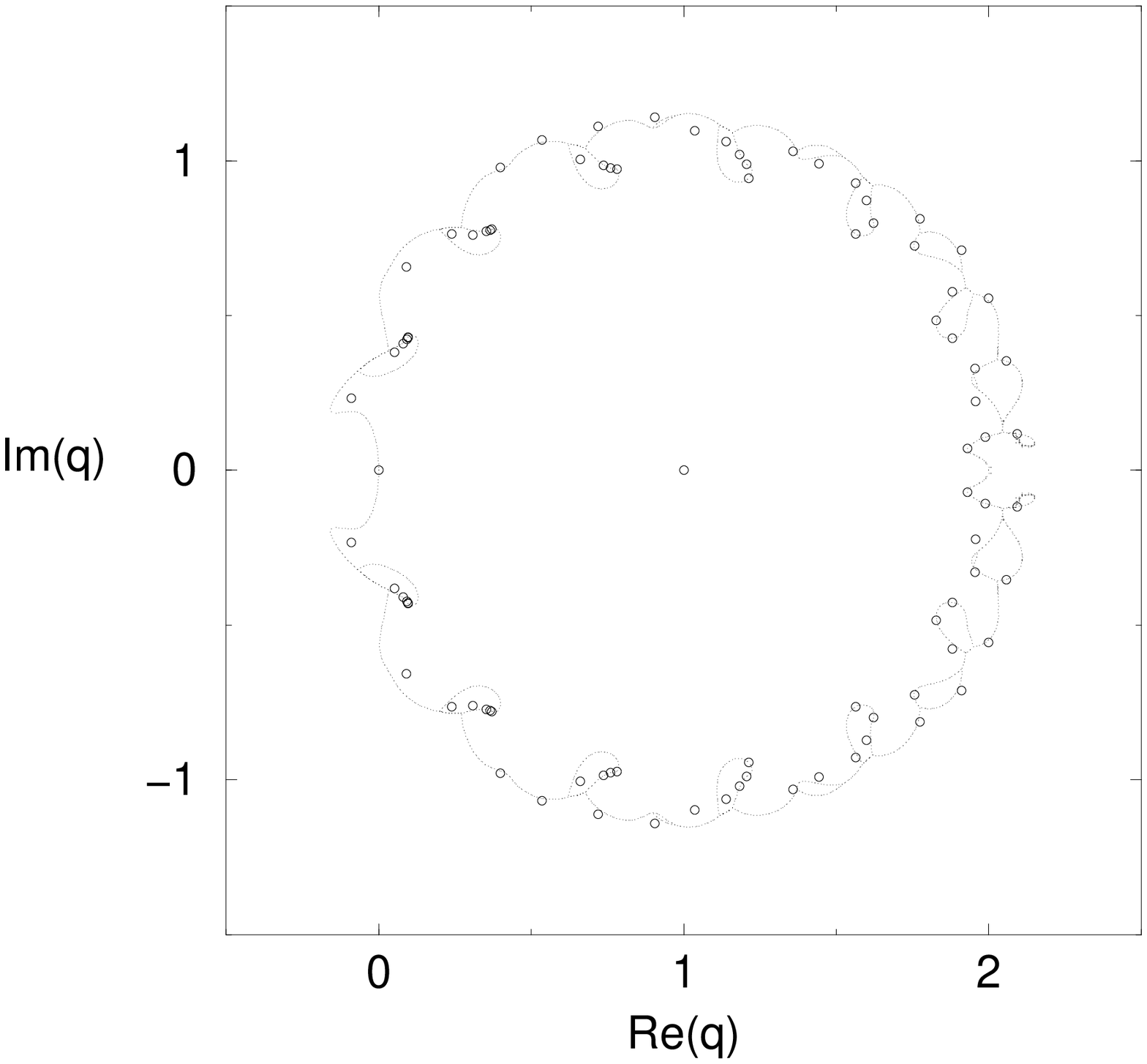}
\end{center}
\caption{\footnotesize{Locus ${\cal B}$ for the $m \to \infty$ limit of the
family of cyclic graphs $X_{15,m}$ and chromatic zeros for the cyclic graph
$X_{15,m}$ with $m=6$ (i.e., $n=90$).}}
\label{E15zeros}
\end{figure}

\begin{figure}[hbtp]
\centering
\leavevmode
\epsfxsize=4.0in
\begin{center}
\leavevmode
\epsffile{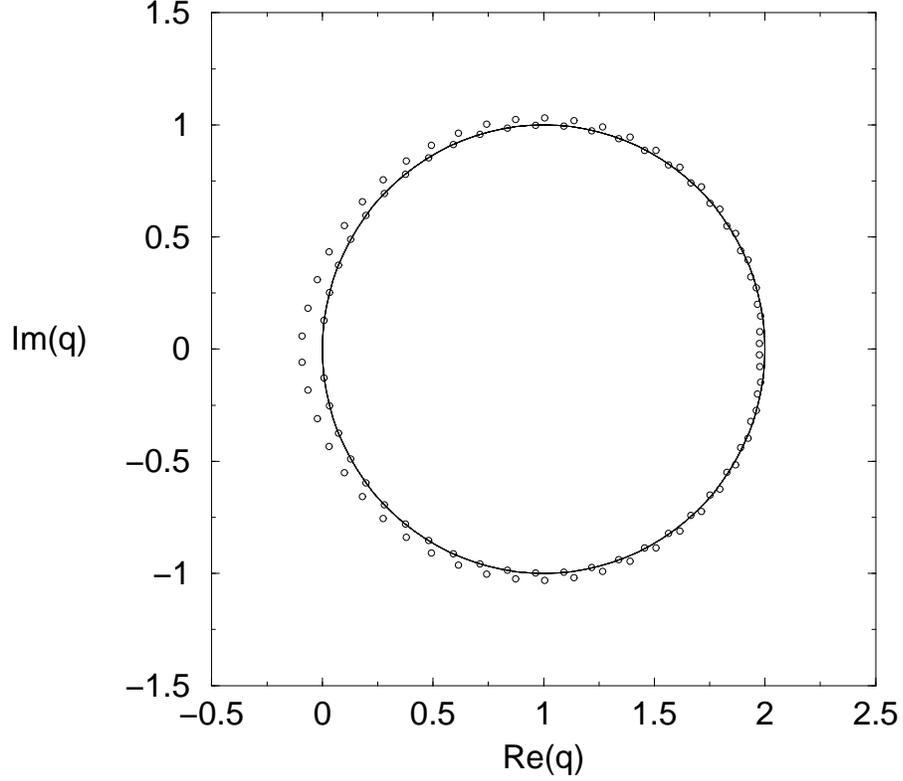}
\end{center}
\caption{\footnotesize{The roots of the discriminant of eq. (\ref{lamE34})
with $k=50$. The solid circle is $|q-1|=1$. }}
\label{E50roots}
\end{figure}

As $k$ increases, more small regions appear on the boundary between
regions $R_1$ and $R_2$. These small regions surround part of the roots of
the discriminant of eq. (\ref{lamE34}), as shown in Fig.  \ref{E7zeros}
for $k=7$ and Fig. \ref{E8zeros} for $k=8$.  In these figures the roots of
the discriminant are indicated with an asterisk, $*$. Note that the locus
${\cal B}$ is compact and approaches the circle $|q-1|=1$ as $k$
increases, as is illustrated in Fig. \ref{E15zeros} for $k=15$.  This is
similar to what was found before for the outer envelope of the locus
${\cal B}$ for the homeomorphic expansions on the horizontal edges of the
cyclic $L_y=2$ strips of the square lattices \cite{nec}.  The roots of the
discriminant of eq.  (\ref{lamE34}) also approach the same circle as $k$
increases, as is illustrated in Fig. \ref{E50roots} for $k=50$.

\section{Family of $Y_{k,m}$  Strips with $(FBC_y,PBC_x)$}

In this section we insert $k-2$ degree-2 vertices on all the vertical
edges of the $L_y=2$ triangular lattice with cyclic boundary conditions. 
Here we find $N_{Y,\lambda}=4$, and

\beq
P(Y_{k,m},FBC_y,PBC_x,q) = \sum_{j=1}^{4} c_{Y,j} (\lambda_{Y_k,j})^m \ ,
\label{pF}
\eeq
where 

\beqs
\lambda_{Y_k,1} & = & (q-1)D_{k+1}(q)-(q-2)D_k(q) \cr\cr
                & = &  D_4(q)D_k(q)+(q-1)(-1)^{(k-1)} \ , 
\label{lamF1} 
\eeqs

The $\lambda_{Y_k,(2,3)}$ are the roots of the quadratic equation,

\beqs
& & \xi^2+\frac{(-1)^k}{q}\Bigl ( 2(2-q)(1-q)^{(k-1)} + q-4 \Bigr )\xi
\cr\cr
& & +\frac{1}{q^2}\Bigl ( (q-3)(q-1)^{(2k-1)} - (q-2)(2q-3)(1-q)^{(k-1)}
      +q^2-3q+3 \Bigr ) =0 \ ,
\label{lamF23}
\eeqs
where the coefficient functions are polynomials for positive integer $k$,
and

\beq
\lambda_{Y_k,4} = D_k(q) \ .
\label{lamF4}
\eeq

The corresponding coefficients are

\beq
c_{Y,1}=1
\label{cF1}
\eeq 

\beq
c_{Y,(2,3)}=c^{(1)}
\label{cF23}
\eeq

and

\beq
c_{Y,4}=c^{(2)} \ .
\label{cF4}
\eeq

\begin{figure}[hbtp]
\centering
\leavevmode
\epsfxsize=4.0in
\begin{center}
\leavevmode
\epsffile{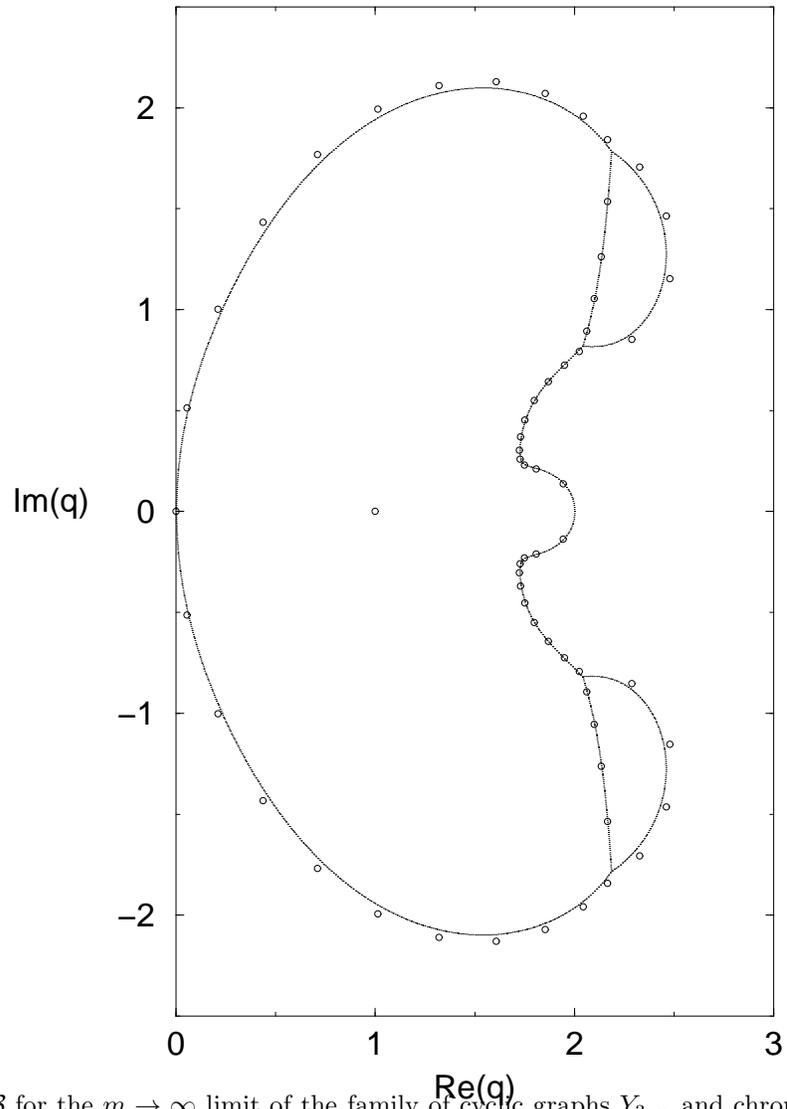}
\end{center}
\caption{\footnotesize{Locus ${\cal B}$ for the $m \to \infty$ limit of the
family of cyclic graphs $Y_{3,m}$ and chromatic zeros for the cyclic graph
$Y_{3,m}$ with $m=20$ (i.e., $n=60$).}}
\label{F3zeros}
\end{figure}

\begin{figure}[hbtp]
\centering
\leavevmode
\epsfxsize=4.0in
\begin{center}
\leavevmode
\epsffile{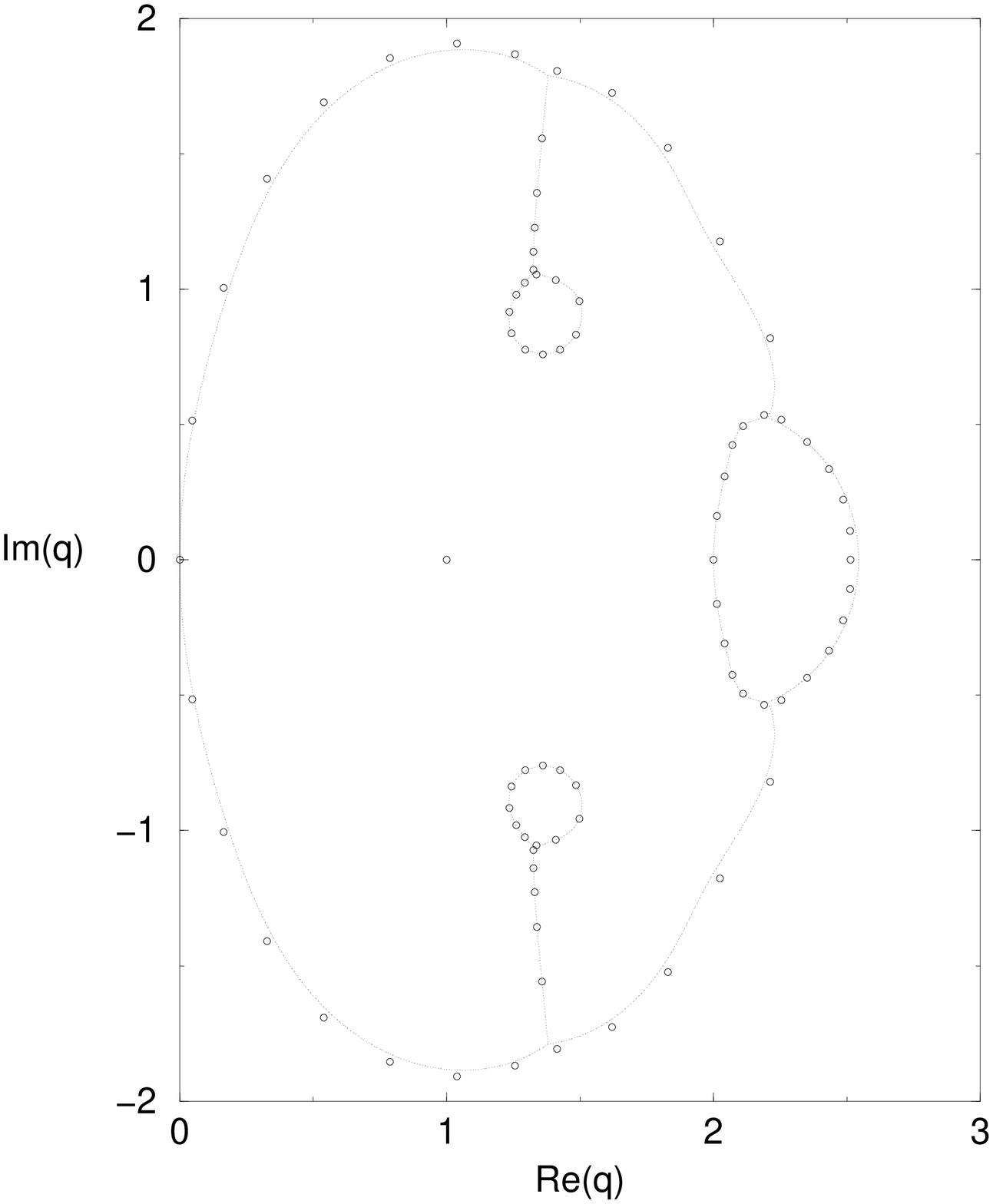}
\end{center}
\caption{\footnotesize{Locus ${\cal B}$ for the $m \to \infty$ limit of the
family of cyclic graphs $Y_{4,m}$ and chromatic zeros for the cyclic graph
$Y_{4,m}$ with $m=20$ (i.e., $n=80$).}}
\label{F4zeros}
\end{figure}

These coefficients are the same as the coefficients for the square and
triangular lattices with $L_y=2$ and the same boundary condition. 
Therefore, the sum of the coefficients is equal to $P(T_2,q)=q(q-1)$ if we
do not consider the zero eigenvalue. The chromatic number is 2 as
$(k,L_x)=(o,e)$, and 3 as $(k,L_x)=(o,o), (e,e), (e,o)$, where $e$ and $o$
denote even and odd values for $k$ and $L_x$. 
The locus ${\cal B}$ for the $m \to \infty$ limit of the cyclic family 
$Y_{k,m}$ and the chromatic zeros for the $L_x=m=20$ cyclic graph 
$Y_{k,m}$ with $k=3, 4$ are shown in Fig.
\ref{F3zeros} and \ref{F4zeros}. 

The locus ${\cal B}$ has support for $Re(q) \ge 0$, and crosses the real
$q$-axis at $q=0,2$ for odd $k$ and $q=0,2,q_c$ for even $k$, where $2 \le q_c
\le 3$. The two main regions for odd $k$ are region $R_1$, which includes the
intervals $q \ge 2$ and $q \le 0$ on the real $q$ axis, and region $R_2$, which
includes the real interval $0 \le q \le 2$. For even $k$, there is another
region $R_3$ which includes the real interval $2 \le q \le q_c$, together with
region $R_1$ including the intervals $q \ge q_c$ and $q \le 0$ on the real $q$
axis. The value of this $q_c$ decreases as $k$ increases. The dominant terms in
regions $R_1$ and $R_2$ are $\lambda_{Y,1}$ in (\ref{lamF1}) and the root of
eq. (\ref{lamF23}). For even $k$, the dominant term in $R_3$ is $\lambda_{Y,4}$
in (\ref{lamF4}).

For $k=3$, there is also a pair of complex-conjugate regions $R_3, R_3^*$
contiguous to $R_2$, and the dominant term in these regions is
$\lambda_{Y_3,3}$. The ${\cal B}(Y_k)$ around $q_c$ again has the form of
a circular arc that bends to the left before reaching a pair of roots of the
discriminant of eq. (\ref{lamF23}).

\begin{figure}[hbtp]
\centering  
\leavevmode
\epsfxsize=4.0in
\begin{center}
\leavevmode
\epsffile{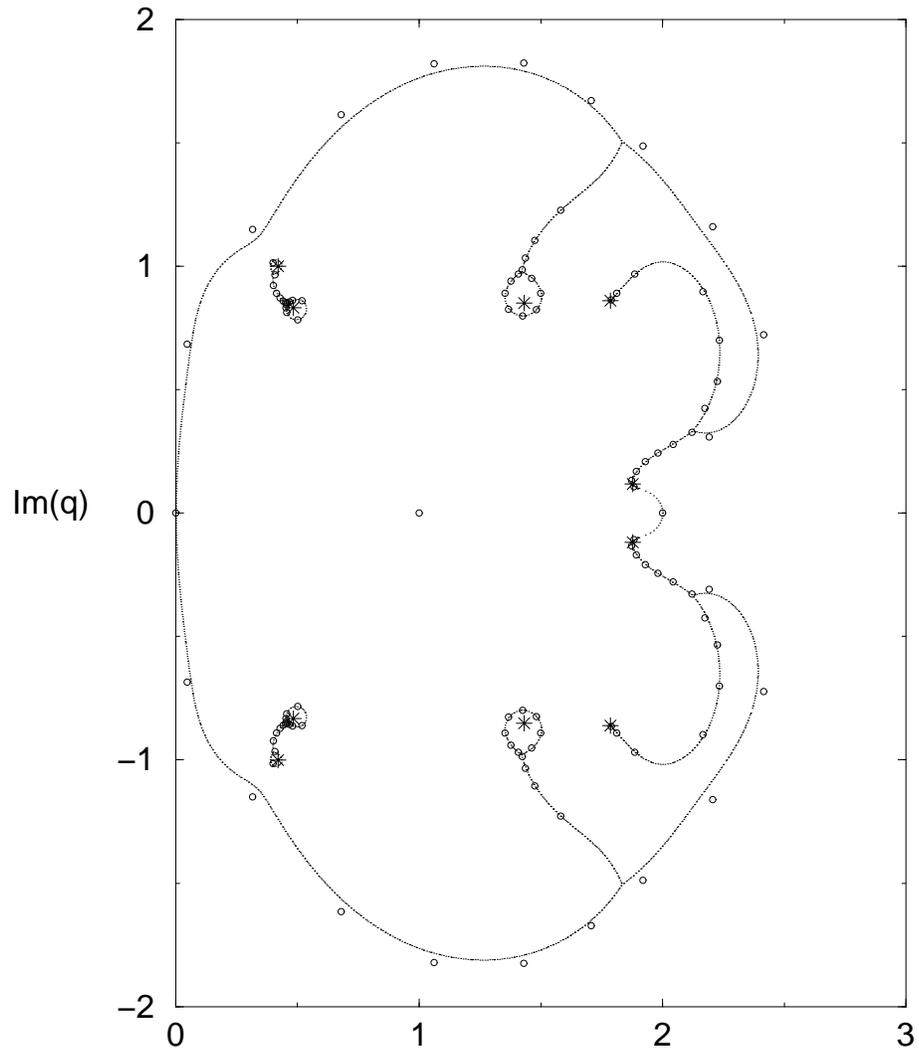}
\end{center}
\caption{\footnotesize{Locus ${\cal B}$ for the $m \to \infty$ limit of the
cyclic family $Y_{7,m}$ and chromatic zeros for the cyclic graph 
$Y_{7,m}$ with $m=15$ (i.e., $n=105$). The roots of the discriminant of
eq. (\ref{lamF23}) are indicated with $*$. }}
\label{F7zeros}
\end{figure}

\begin{figure}[hbtp]
\centering  
\leavevmode
\epsfxsize=4.0in
\begin{center}
\leavevmode
\epsffile{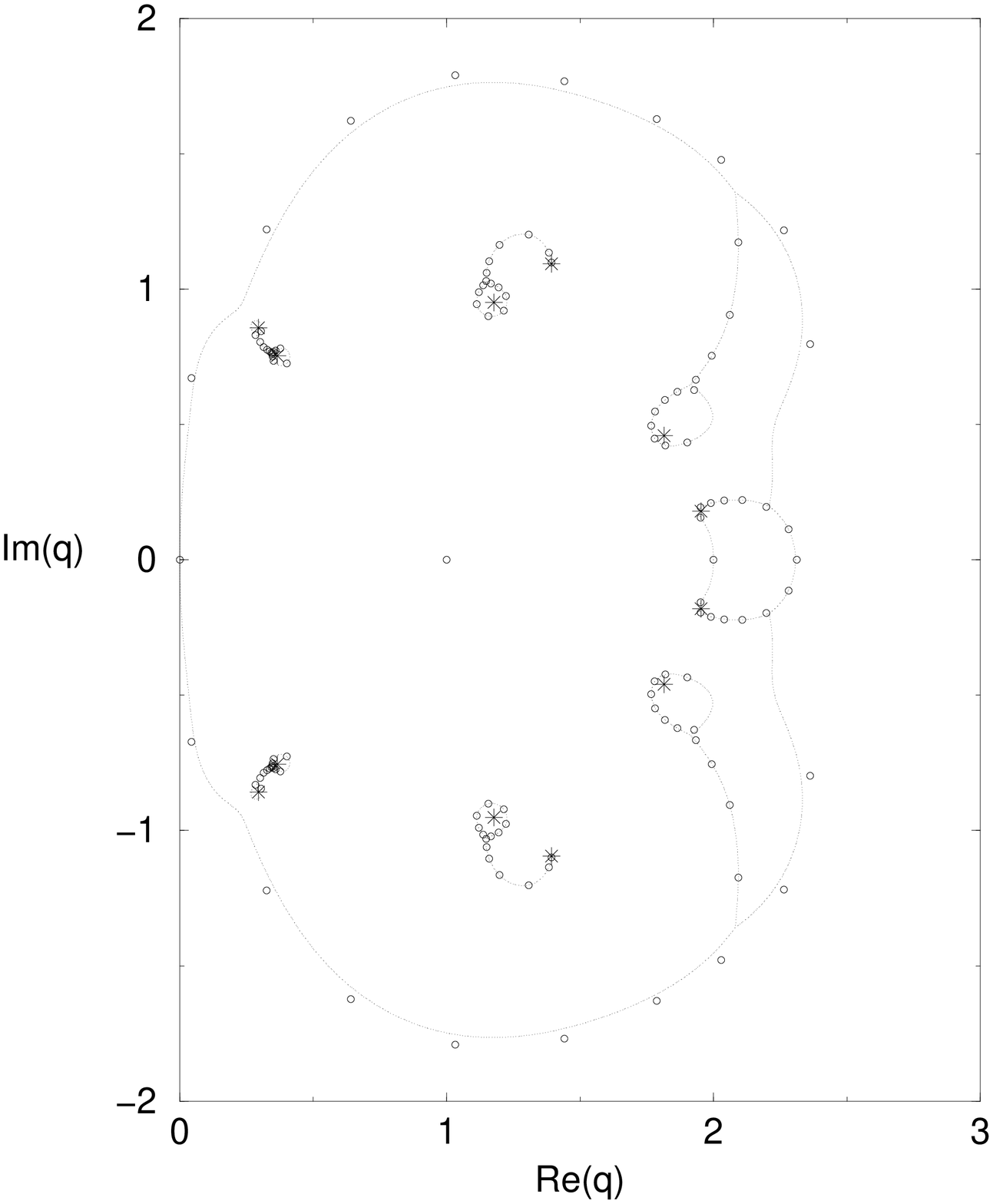}
\end{center}
\caption{\footnotesize{Locus ${\cal B}$ for the $m \to \infty$ limit of the
cyclic family $Y_{8,m}$ and chromatic zeros for the cyclic graph 
$Y_{8,m}$ with $m=15$ (i.e., $n=120$). The roots of the discriminant of
eq. (\ref{lamF23}) are indicated with $*$. }}
\label{F8zeros}
\end{figure}

\begin{figure}[hbtp]
\centering  
\leavevmode
\epsfxsize=4.0in
\begin{center}
\leavevmode
\epsffile{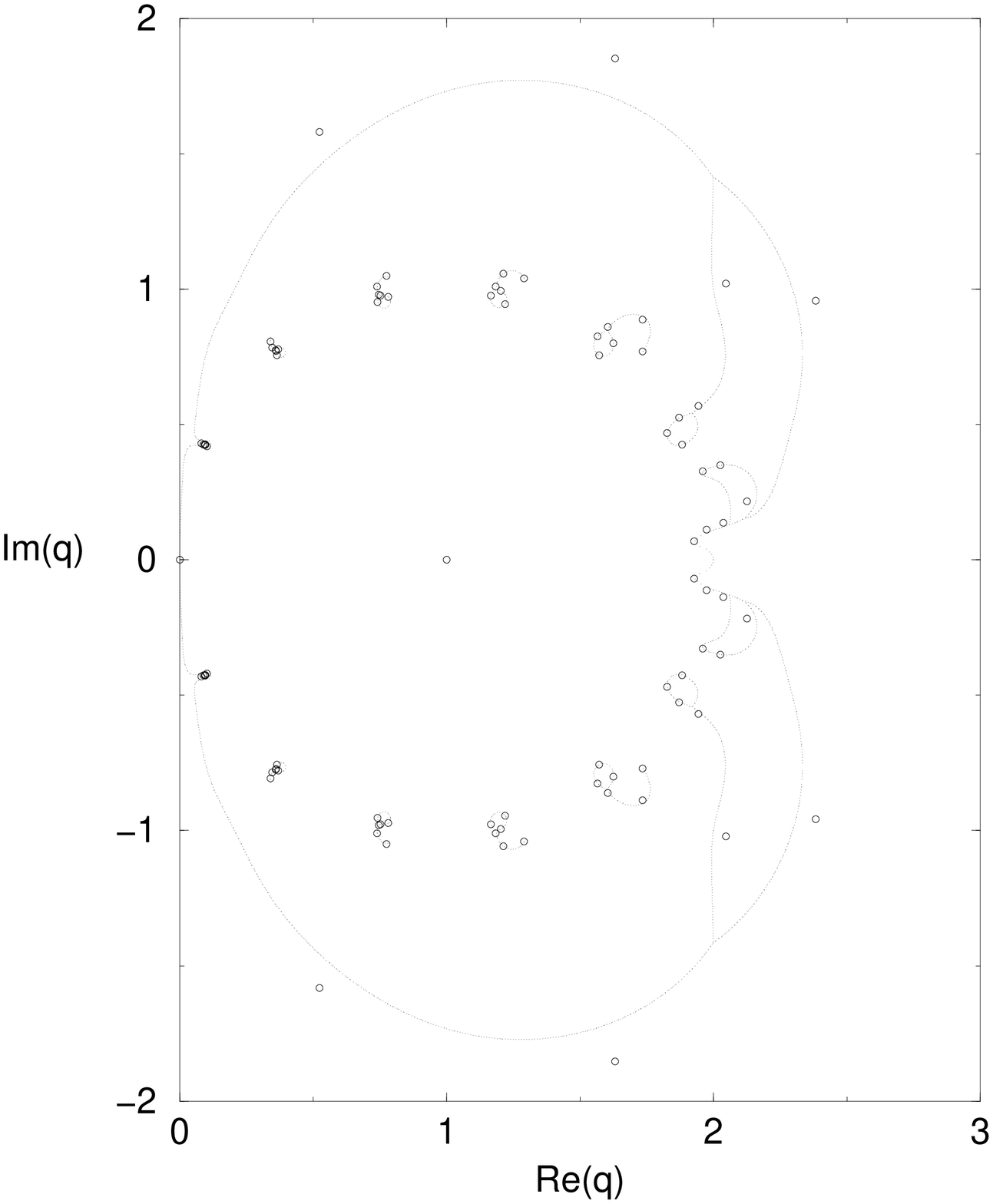}
\end{center}
\caption{\footnotesize{Locus ${\cal B}$ for the $m \to \infty$ limit of the
cyclic family $Y_{15,m}$ and chromatic zeros for the cyclic graph
$Y_{15,m}$ with $m=6$ (i.e., $n=90$).}}
\label{F15zeros}
\end{figure}

\begin{figure}[hbtp]
\centering  
\leavevmode
\epsfxsize=4.0in
\begin{center}
\leavevmode
\epsffile{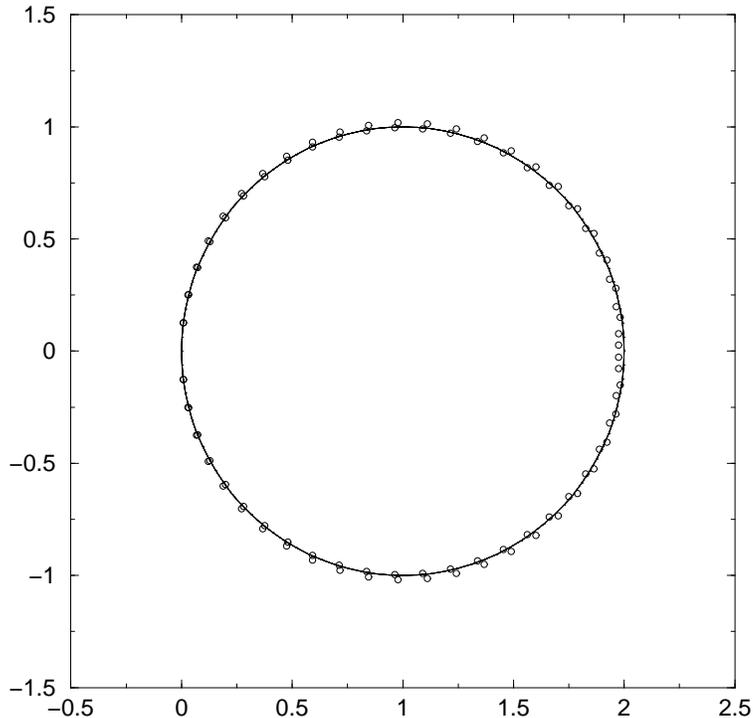}
\end{center}
\caption{\footnotesize{The roots of the discriminant of eq. (\ref{lamF23})
with $k=50$. The solid circle is $|q-1|=1$. }}
\label{F50roots}
\end{figure}

For $k \ge 4$, there are more regions that appear inside $R_2$; in these
regions, the dominant term is $\lambda_{Y_k,1}$.  These internal regions are
connected to an arc, and the free end of the most right-hand arc is connected
to the boundary between regions $R_1$ and $R_2$. The most left-land arc has two
small regions on both of its ended if the small region is not connected to the
boundary between regions $R_1$ and $R_2$.  The free ends of these arcs are
roots of the discriminant of eq. (\ref{lamF23}), and the small regions they are
connected to also surround the roots of the discriminant. This is illustrated
in Fig. \ref{F7zeros} and Fig. \ref{F8zeros}, where the roots of the
discriminant are indicated with $*$. These internal regions approach the 
circle $|q-1|=1$ as $k$ increases, as is indicated in Fig.
\ref{F15zeros} for $k=15$.  The roots of the discriminant of eq.
(\ref{lamF23}) also approach the same circle as $k$ increases, as is 
illustrated in Fig. \ref{F50roots} for $k=50$. 

\section{Comparative Discussion}

We have found some interesting features of the locus ${\cal B}$ for these 
strips with $(FBC_y, PBC_x)=$ cyclic boundary condition:

\begin{enumerate}

\item

For the families of the lattice strips that we have studied so far, as long as
the boundary conditions are cyclic, i.e., each repeated vertex set forms a tree
graph $(T_{L_y})$ and the last vertex set is connected to the first vertex set
to have periodic boundary conditions in the longitudinal direction, the
coefficients $c_{G_s,j}(q)$'s that enter into the expressions for the chromatic
polynomial (\ref{pgsum}) are always in the form of Chebyshev polynomial of the
second kind given in eq. $(\ref{cd})$. This property was found to be true for
square and triangular lattices with cyclic boundary condition in
\cite{cf}. Note that, while this statement applies also to certain classes of
nonplanar strip graphs, such as the M\"obius strips of the square lattice, it
does not apply to the nonplanar strip graphs that we have analyzed in  
\cite{k}; for these, we showed that the coefficients have different forms and 
are similar to the coefficients of the strips with doubly periodic boundary
condition (torus) \cite{t,tk,tor4}.

\item

In general, the sum of the coefficients $C(G_s)$ in eq. ($\ref{cgsum}$) is the
total dimension of the space of coloring configurations and should be equal to
$P(T_{L_y})=q(q-1)^{L_y-1}$ for the lattice strips with $(FBC_y, PBC_x)$
boundary conditions. Lattice $Q_m$ does satisfies this relation, as shown in
eq. (\ref{cbsum}). For other lattice strips studied here, the sum of the
coefficients without taking account the coefficient of the zero eigenvalue are
not $P(T_{L_y})$, but this can be understood easily. Consider two adjacent
vertex sets of lattice $X_{k,m}$ with $k \ge 3$ and apply the coloring matrix
from one vertex set to another. Effectively, only the two end vertices are
connected and should be considered for coloring.  Since they are separated by
other vertices in between, the total number of proper colorings of these
vertices is $q^2$. A similar argument applies to the lattice strip $V_m$ in
Fig. \ref{illustration} (d), and the lattice strip $S_m$ in Fig.
\ref{deformation} (a). For the lattice strip $O_m$,
effectively only two connected
vertices (forming a $T_2$) and a separated vertex in each vertex set are
connected to another vertex set; therefore the sum of the coefficients is
$q^2(q-1)$.  With the homeomorphic expansions on the vertical edges of the
cyclic $L_y=2$ strips of the triangular lattice, $Y_{k,m}$ does not satisfy the
condition C given in \cite{lse0004}. However, it is equivalent to the
homeomorphic expansions on the diagonal edges of the same lattice strip, as
shown in Fig. \ref{deformation} (b), so the sum of the coefficients is still
$q(q-1)$.

\vspace*{1cm}
\begin{picture}(100,20)
\multiput(0,0)(10,0){5}{\circle*{2}}
\multiput(0,10)(10,0){5}{\circle*{2}}
\multiput(0,20)(10,0){5}{\circle*{2}}
\multiput(0,0)(10,0){5}{\line(0,1){20}}
\multiput(0,0)(0,20){2}{\line(1,0){40}}
\multiput(0,0)(10,0){4}{\line(1,1){10}}
\multiput(0,10)(10,0){4}{\line(1,1){10}}
\put(-2,-2){\makebox(0,0){9}}
\put(8,-2){\makebox(0,0){10}}
\put(18,-2){\makebox(0,0){11}}
\put(28,-2){\makebox(0,0){12}}
\put(38,-2){\makebox(0,0){9}}
\put(-2,12){\makebox(0,0){5}}
\put(8,12){\makebox(0,0){6}}
\put(18,12){\makebox(0,0){7}}
\put(28,12){\makebox(0,0){8}}
\put(38,12){\makebox(0,0){5}}
\put(-2,22){\makebox(0,0){1}}
\put(8,22){\makebox(0,0){2}} 
\put(18,22){\makebox(0,0){3}} 
\put(28,22){\makebox(0,0){4}} 
\put(38,22){\makebox(0,0){1}}
\put(50,10){\makebox(0,0){=}}
\multiput(60,0)(10,0){5}{\circle*{2}}
\multiput(60,10)(10,0){5}{\circle*{2}}
\multiput(60,20)(10,0){5}{\circle*{2}}
\multiput(60,0)(10,0){5}{\line(0,1){20}}
\multiput(60,0)(0,20){2}{\line(1,0){40}}
\multiput(60,10)(10,0){4}{\line(1,-1){10}}
\multiput(60,10)(10,0){4}{\line(1,1){10}}
\put(58,-2){\makebox(0,0){12}}
\put(68,-2){\makebox(0,0){9}}
\put(78,-2){\makebox(0,0){10}}
\put(88,-2){\makebox(0,0){11}}
\put(98,-2){\makebox(0,0){12}}
\put(58,12){\makebox(0,0){5}}
\put(68,12){\makebox(0,0){6}}
\put(78,12){\makebox(0,0){7}}
\put(88,12){\makebox(0,0){8}}
\put(98,12){\makebox(0,0){5}}
\put(58,22){\makebox(0,0){1}}
\put(68,22){\makebox(0,0){2}} 
\put(78,22){\makebox(0,0){3}} 
\put(88,22){\makebox(0,0){4}} 
\put(98,22){\makebox(0,0){1}}
\put(50,-8){\makebox(0,0){(a)}}
\end{picture}
\vspace*{3cm}

\begin{picture}(100,20)
\multiput(0,0)(10,0){5}{\circle*{2}}
\multiput(0,10)(10,0){5}{\circle*{2}}
\multiput(0,20)(10,0){5}{\circle*{2}}
\multiput(0,0)(10,0){5}{\line(0,1){20}}
\multiput(0,0)(0,20){2}{\line(1,0){40}}
\multiput(0,0)(10,0){4}{\line(1,2){10}}
\put(-2,-2){\makebox(0,0){9}}
\put(8,-2){\makebox(0,0){10}}
\put(18,-2){\makebox(0,0){11}}
\put(28,-2){\makebox(0,0){12}}
\put(38,-2){\makebox(0,0){9}}
\put(-2,12){\makebox(0,0){5}}
\put(8,12){\makebox(0,0){6}}
\put(18,12){\makebox(0,0){7}}
\put(28,12){\makebox(0,0){8}}
\put(38,12){\makebox(0,0){5}}
\put(-2,22){\makebox(0,0){1}}
\put(8,22){\makebox(0,0){2}} 
\put(18,22){\makebox(0,0){3}} 
\put(28,22){\makebox(0,0){4}} 
\put(38,22){\makebox(0,0){1}}
\put(50,10){\makebox(0,0){=}}
\multiput(60,0)(10,0){5}{\circle*{2}}
\multiput(65,10)(10,0){4}{\circle*{2}}
\multiput(60,20)(10,0){5}{\circle*{2}}
\multiput(60,0)(10,0){5}{\line(0,1){20}}
\multiput(60,0)(0,20){2}{\line(1,0){40}}
\multiput(60,20)(10,0){4}{\line(1,-2){10}}
\put(58,-2){\makebox(0,0){12}}
\put(68,-2){\makebox(0,0){9}}
\put(78,-2){\makebox(0,0){10}}
\put(88,-2){\makebox(0,0){11}}
\put(98,-2){\makebox(0,0){12}}
\put(63,12){\makebox(0,0){5}}
\put(73,12){\makebox(0,0){6}}
\put(83,12){\makebox(0,0){7}}
\put(93,12){\makebox(0,0){8}}
\put(58,22){\makebox(0,0){1}}
\put(68,22){\makebox(0,0){2}} 
\put(78,22){\makebox(0,0){3}} 
\put(88,22){\makebox(0,0){4}} 
\put(98,22){\makebox(0,0){1}}
\put(50,-8){\makebox(0,0){(b)}}
\end{picture}
\vspace*{1cm}

\begin{figure}[hbtp]
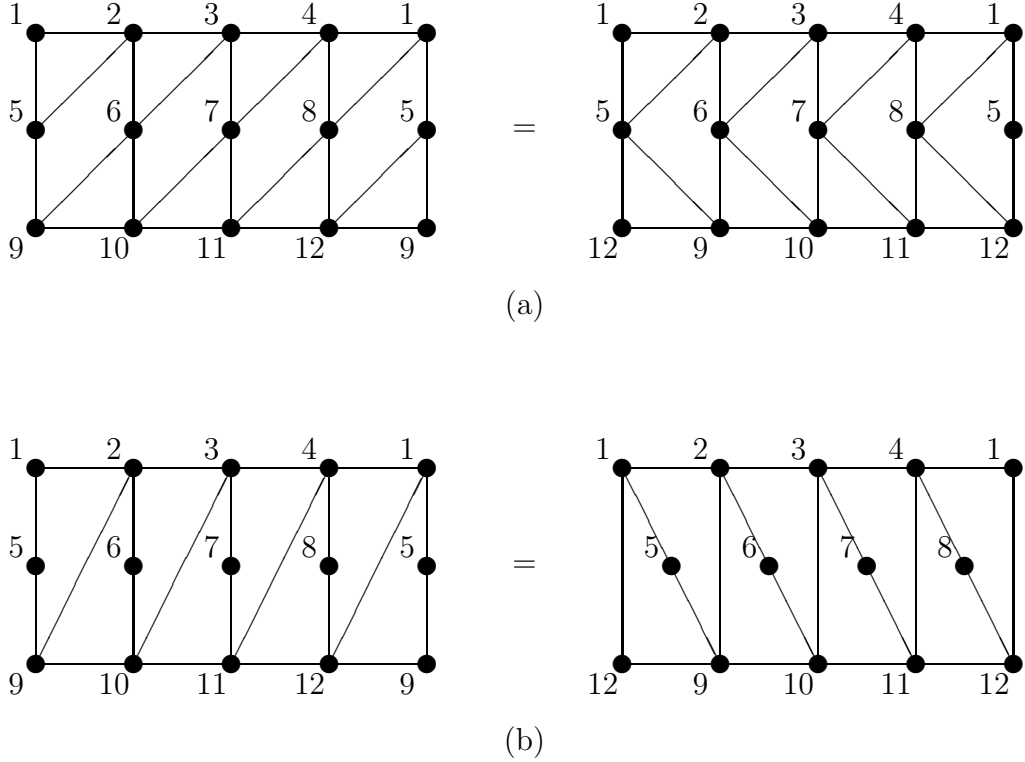
 

\caption{\footnotesize{Illustrative deformation for the strip graphs of
(a) $S_m$ lattice with $L_y=3$ and (b) $Y_{k,m}$ lattice with $k=3$,
length $L_x=m=4$ and $(FBC_y,PBC_x)=$ cyclic boundary conditions.}} 

\label{deformation} 
\end{figure}

\item

The loci ${\cal B}$'s for the infinite-length limits of the strips studied here
with global circuits in the longitudinal direction cross the real $q$-axis at
$q=0,2$ as before.  The feature of the cyclic strips that there is a
correlation between the coefficient $c_{G_s,j}$ of the respective dominant
$\lambda_{G_s,j}$'s in regions that include intervals of the real axis
discussed in \cite{t} is found to be true again. The $c_{G_s,j}$'s of the
dominant $\lambda_{G_s,j}$'s in (1) region $R_1$ containing the intervals $q
\ge q_c$ and $q \le 0$, (2) region containing the intervals $0 < q < 2$, and
(3) region containing the interval $2 < q < q_s$, where $q_s$ may not be $q_c$,
are $c^{(0)}=1, c^{(1)}$, and $c^{(2)}$, respectively.

\item 

In previous calculations of ${\cal B}$ for the zero-temperature Potts
antiferromagnet on infinite-length limits of strip graphs with global
circuits (i.e., $PBC_x$) \cite{w,wcy,nec,pg,pm,tk,s4,t}, the respective
loci did not involve arc endpoints.  Here we find such arcs with endpoints
such that one end is connected to the boundary of regions for the
infinite-length limits of the families $S$, $V$, and $Y_k$ with $k \ge 7$,
and arcs connecting two regions for lattices $V$, $X_{odd}$ and $Y_k$ with
$k \ge 4$. In Table \ref{proptable} we list various properties of these
lattices presented in this work. 

\item

Although some loci ${\cal B}$ calculated for the infinite-length limits of
families of graphs with global circuits involved disconnected pieces (e.g., 
\cite{wa2,nec}), here we find a series of such disconnected components, with a
remarkably intricate structure, as is evident in Figs. \ref{F7zeros}, 
\ref{F8zeros}, and \ref{F15zeros}.

\begin{table}
\caption{\footnotesize{Properties of $P$, $W$, and ${\cal B}$ for strip
graphs $G_s$ of lattices studied here. The properties apply for a given
strip of type $G_s$ of size $L_y \times L_x$; some apply for arbitrary
$L_x$, such as $N_\lambda$, while others apply for the infinite-length
limit, such as the properties of the locus ${\cal B}$.  The boundary
conditions is ($FBC_y,PBC_x$) = cyclic.  The column denoted eqs. 
describes the numbers and degrees of the algebraic equations giving the
$\lambda_{G_s,j}$; for example, $\{2(1),2(2),1(4)\}$ indicates that there
are 2 linear equations, 2 quadratic equations and one quartic equation.
The column denoted BCR lists the points at which ${\cal B}$ crosses the
real $q$ axis; the largest of these is $q_c$ for the given family $G_s$. 
The notation ``int;$q_1;q_c$'' refers to cases where ${\cal B}$ contains a
real interval, there is a crossing at $q_1$, and the right-hand endpoint
of the interval is $q_c$. $q_c$ for lattice $Y_{k,m}$ with even $k$
depends on the value of $k$, and is not shown here.  Column labeled ``SN''
refers to whether ${\cal B}$ has \underline{s}upport for
\underline{n}egative $Re(q)$, indicated as yes (y) or no (n).}}
\begin{center} 
\begin{tabular}{|c|c|c|c|c|c|} 
\hline\hline $G_s$ & $L_y$ & $N_\lambda$ & eqs. & BCR & SN \\ \hline\hline
O & 4 & 14 & \{1(1)1(3)1(4)1(6)\} & 0,\ 2, \ 2.638 & y \\ \hline 
Q & 3 & 10 & \{2(1)2(2)1(4)\}     & 0, \ 2, \ 3     & n \\ \hline 
S & 3 & 6  & \{2(1)2(2)\}         & 0, \ 2, \ 2.703 & n \\ \hline 
V & 3 & 6  & \{1(1)1(2)1(3)\}     & 0, \ 2, \ 2.341, \ int: \ 2.5; \ 3 & 
n \\ \hline 
X & k & 6  & \{2(1)2(2)\}         & 0, \ 2       & y \\ \hline 
Y & k & 4  & \{2(1)1(2)\}         & 0, \ 2,      $q_c(k_{even})$ & n \\
\hline\hline
\end{tabular}
\end{center} 
\label{proptable} 
\end{table}

\end{enumerate}

\section{Conclusions}

In this paper we have presented exact calculations of the zero-temperature
$q$-state Potts antiferromagnet partition function (equivalently, chromatic
polynomial $P$), for various lattice strips of fixed width $L_y$ and
arbitrarily great length $L_x$ with cyclic boundary conditions $(FBC_y,PBC_x)$.
The coefficient $c_{G,j}$ of degree $d$ in $q$ is
$c^{(d)}=U_{2d}(\frac{\sqrt{q}}{2})$, where $U_n(x)$ is the Chebyshev
polynomial of the second kind. We explain the sums of the coefficients for the
respective families.  We also find a number of interesting and novel features
of the singular locus ${\cal B}$.  The results herein are thus of interest both
from the viewpoint of exact results in statistical mechanics, graph theory, and
algebraic geometry.

\section{Appendix: Family of $Z_m$ Strips with $(PBC_y,PBC_x)$}

\begin{figure}[hbtp]
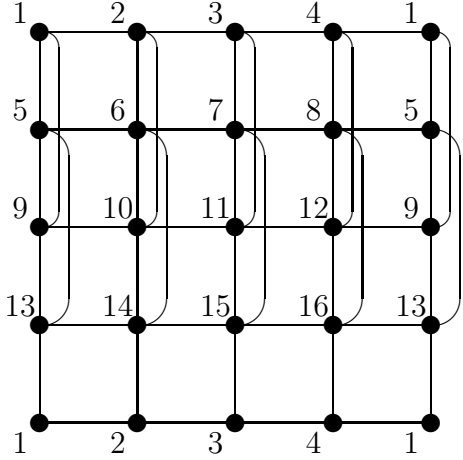
 

\caption{\footnotesize{Illustrative strip graphs of $Z_m$ lattice with
$L_y=4$ with length $L_x=m=4$ and $(PBC_y,PBC_x)$ = torus boundary
condition. Vertices are indicated with $\bullet$ (points where edges cross
without a symbol $\bullet$ are not vertices.) }}

\label{illustrationH} 
\end{figure}

In this section we consider the $4 \times m$ strips of the square lattice
with $(PBC_y,PBC_x)$ boundary conditions (torus), but with two diagonal
edges in each basic vertex set , i.e., instead of $L_x$ set of tree
graphs, we have $L_x$ set of complete graph\footnote{$K_n$ denotes
the complete graph, i.e. the graph with $n$ vertices such that each vertex
is connected by an edge to all of the other vertices.} $K_4$ (tetrahedron) to
begin with and $e=\{11,22,33,44\}$. The lattice with length $L_x=m$ will
be denoted as $Z_m$, and an illustrative example with $m=4$ is displayed
in Fig. \ref{illustrationH}. 

\vspace*{1cm}
\begin{picture}(40,40)
\multiput(0,0)(10,0){5}{\circle*{2}}
\multiput(0,10)(10,0){5}{\circle*{2}}
\multiput(0,20)(10,0){5}{\circle*{2}}
\multiput(0,30)(10,0){5}{\circle*{2}}
\multiput(0,40)(10,0){5}{\circle*{2}}
\multiput(0,0)(10,0){5}{\line(0,1){40}}
\multiput(0,0)(0,10){5}{\line(1,0){40}}
\multiput(0,20)(10,0){5}{\oval(6,20)[r]}
\multiput(0,30)(10,0){5}{\oval(4,20)[r]}
\put(-2,-2){\makebox(0,0){1}}
\put(8,-2){\makebox(0,0){2}}
\put(18,-2){\makebox(0,0){3}}
\put(28,-2){\makebox(0,0){4}}
\put(38,-2){\makebox(0,0){1}}
\put(-2,12){\makebox(0,0){13}}
\put(8,12){\makebox(0,0){14}}
\put(18,12){\makebox(0,0){15}}
\put(28,12){\makebox(0,0){16}}
\put(38,12){\makebox(0,0){13}}
\put(-2,22){\makebox(0,0){9}}
\put(8,22){\makebox(0,0){10}} 
\put(18,22){\makebox(0,0){11}} 
\put(28,22){\makebox(0,0){12}} 
\put(38,22){\makebox(0,0){9}}
\put(-2,32){\makebox(0,0){5}}
\put(8,32){\makebox(0,0){6}}
\put(18,32){\makebox(0,0){7}}
\put(28,32){\makebox(0,0){8}}
\put(38,32){\makebox(0,0){5}}
\put(-2,42){\makebox(0,0){1}}
\put(8,42){\makebox(0,0){2}} 
\put(18,42){\makebox(0,0){3}} 
\put(28,42){\makebox(0,0){4}} 
\put(38,42){\makebox(0,0){1}}
\end{picture}
\vspace*{1cm}

In \cite{lse0004}, the lattice with $L_x$ set of $K_{L_y}$ for arbitrary
$L_y$ is called the bracelet strip, and the $\lambda$'s with coefficients 1,
$O(q)$, and $O(q^2)$ are given.  For $L_y=2$, it is the ladder graph with
vertex set $T_2=K_2$. For lattice strip with $L_y=3$, which has vertex set
$C_3=K_3$, the chromatic polynomial was given in \cite{tk}.  Here we give
the full set of $\lambda$'s for the $L_y=4$ strip. We find
$N_{Z,\lambda}=16$ and

\beq
P(Z_m,PBC_y,PBC_x,q) = \sum_{j=1}^{16} c_{Z,j} (\lambda_{Z,j})^m \ ,
\label{pH}
\eeq
where all $\lambda_{Z,j}$'s for $1 \le j \le 16$ are polynomials:

\beq
\lambda_{Z,1}=q^4-10q^3+41q^2-84q+73
\label{lamH1}
\eeq

\beq
\lambda_{Z,2}=-(q^3-12q^2+50q-73)
\label{lamH2}
\eeq

\beq
\lambda_{Z,3}=-(q-3)(q^2-5q+7)
\label{lamH3}
\eeq

\beq
\lambda_{Z,4}=q^2-9q+21
\label{lamH4}
\eeq

\beq
\lambda_{Z,5}=q^2-5q+5
\label{lamH5}
\eeq

\beq
\lambda_{Z,6}=q^2-11q+31
\label{lamH6}
\eeq

\beq
\lambda_{Z,7}=q^2-5q+7
\label{lamH7}
\eeq

\beq
\lambda_{Z,8}=q^2-7q+11
\label{lamH8}
\eeq

\beq
\lambda_{Z,9}=1-q
\label{lamH9}
\eeq

\beq
\lambda_{Z,10}=2-q
\label{lamH10}
\eeq

\beq
\lambda_{Z,11}=3-q
\label{lamH11}
\eeq

\beq
\lambda_{Z,12}=4-q
\label{lamH12}
\eeq

\beq
\lambda_{Z,13}=5-q
\label{lamH13}
\eeq

\beq
\lambda_{Z,14}=6-q
\label{lamH14}
\eeq

\beq
\lambda_{Z,15}=7-q
\label{lamH15}
\eeq

and $\lambda_{Z,16}=1$. The corresponding coefficients are
 
\beq
c_{Z,1}=1
\label{cH1}
\eeq

\beq  
c_{Z,2}=\frac13c_{Z,3}=q-1
\label{cH23}
\eeq

\beq  
c_{Z,4}=c_{Z,5}=\frac{3}{2}(q-1)(q-2)
\label{cH45}
\eeq

\beq  
c_{Z,6}=\frac{1}{2}c_{Z,7}=\frac{1}{3}c_{Z,8}=\frac{1}{2}q(q-3)
\label{cH678}
\eeq

\beq  
c_{Z,9}=\frac{1}{3}c_{Z,13}=\frac{1}{6}(q-1)(q-2)(q-3)
\label{cH913}
\eeq

\beq  
c_{Z,10}=c_{Z,14}=\frac{3}{2}c_{Z,12}=q(q-2)(q-4)
\label{cH101412}
\eeq

\beq  
c_{Z,15}=\frac{1}{3}c_{Z,11}=\frac{1}{6}q(q-1)(q-5)
\label{cH1511}
\eeq

and 

\beq  
c_{Z,16}=q^4-10q^3+29q^2-24q+1 \ .
\label{cH16}
\eeq

The sum of all of the coefficients is equal to $P(C_4,q)=q(q-1)D_4(q)$. 
The chromatic number is $\chi(Z) = 4$.

\begin{figure}[hbtp]
\centering
\leavevmode
\epsfxsize=4.0in
\begin{center}
\leavevmode
\epsffile{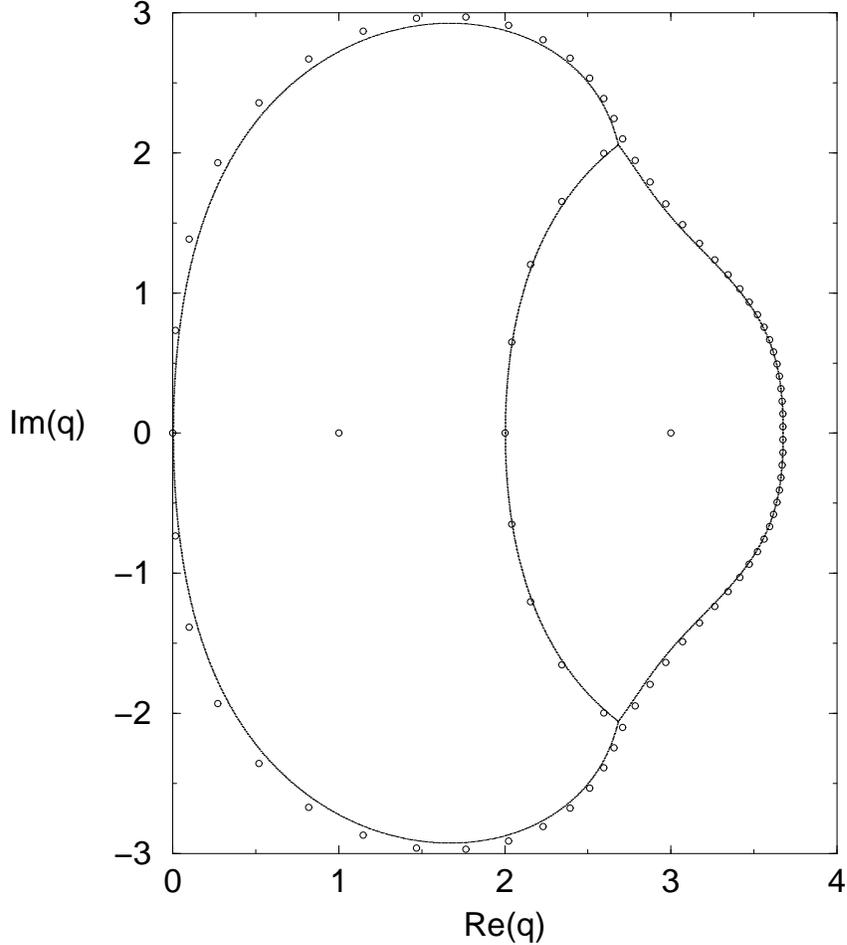}
\end{center}
\caption{\footnotesize{Locus ${\cal B}$ for the $m \to \infty$ limit of the
family $Z_m$ with toroidal boundary conditions and chromatic zeros for 
$Z_m$ with $m=20$ (i.e., $n=80$).}}
\label{Hzeros}
\end{figure}

The locus ${\cal B}$ and chromatic zeros for the $L_x=m=20$ cyclic graph of
the $Z_m$ lattice are shown in Fig. \ref{Hzeros}.  The locus ${\cal B}$
crosses the real $q$-axis at $q=0,2$ and $q_c$, where

\beq
q_c(Z,PBC_y,PBC_x) \simeq 3.673593 \ .
\label{qcH}
\eeq

The locus ${\cal B}$ has support for $Re(q) \ge 0$, and separates the $q$
plane into three regions.  The outermost one, region $R_1$, extends to
infinite $|q|$ and includes the intervals $q \ge q_c$ and $q \le 0$ on the
real $q$ axis.  Region $R_2$ includes the real interval $2 \le q \le q_c$,
while region $R_3$ includes the real interval $0 \le q \le 2$.  In regions
$R_i$, $1 \le i \le 3$, the dominant terms are $\lambda_{Z,1}$,
$\lambda_{Z,6}$, and $\lambda_{Z,2}$, respectively. Thus, the $q_c(Z)$
given in (\ref{qcH}) is the degeneracy between $|\lambda_{Z,1}|$ and
$|\lambda_{Z,6}|$, and is the real solution of $q^3-9q^2+31q-42$.

\begin{figure}[hbtp]
\centering
\leavevmode
\epsfxsize=4.0in
\begin{center}
\leavevmode
\epsffile{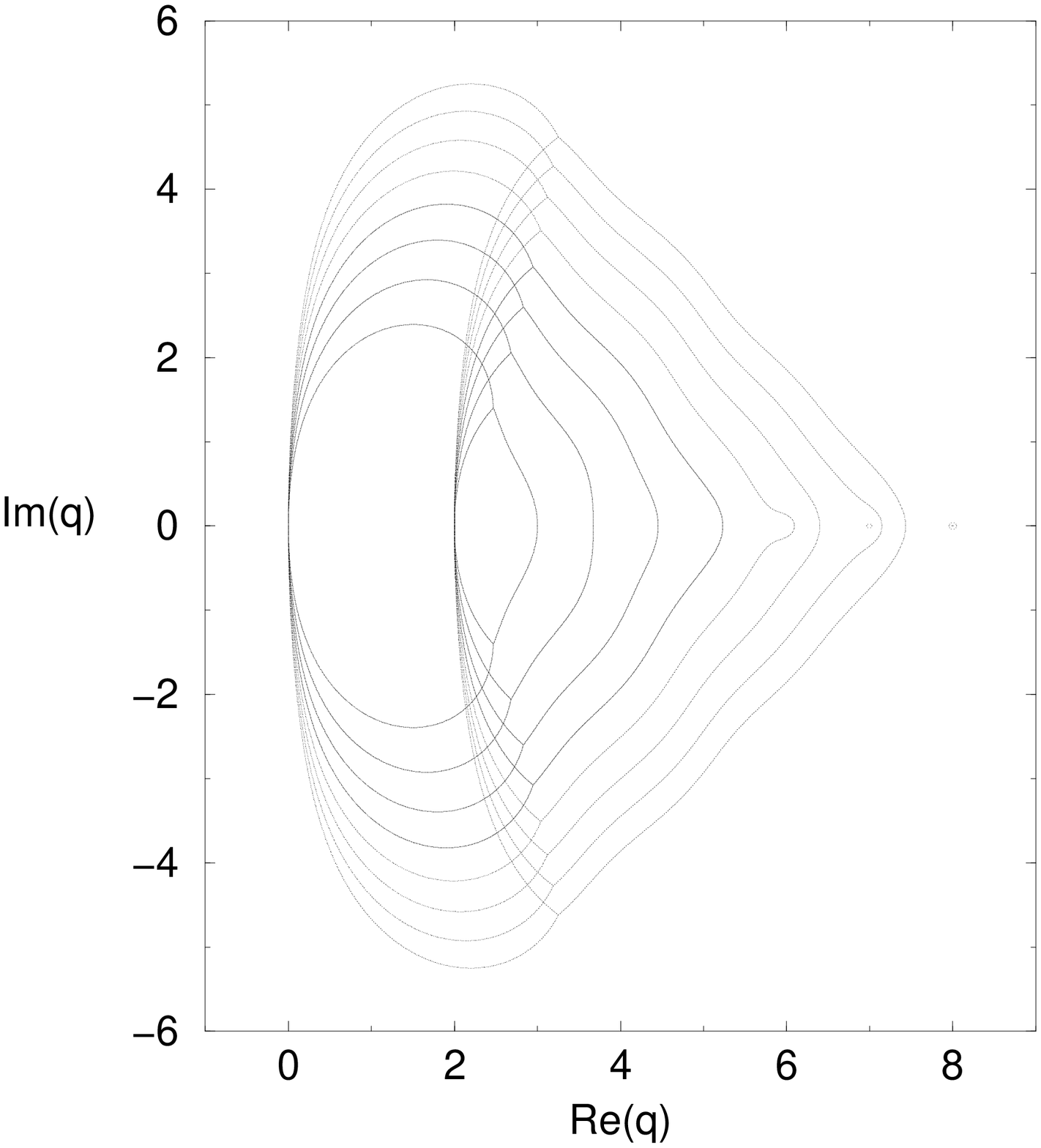}
\end{center}
\caption{\footnotesize{Loci ${\cal B}$'s for the infinite-length limit of 
the square lattice strip with
$K_{L_y}$ as the basic subgraph vertex set and $(PBC_x, PBC_y)$= torus boundary
conditions.  In order of the sizes of the main envelopes from small to
large, the ${\cal B}$'s are for $3 \le L_y \le 10$.}}
\label{sqkpxpy}
\end{figure}

The locus ${\cal B}$ has three regions and crosses the real $q$-axis at
$q=0,2$, and $q_c$, which is quite similar to the ${\cal B}$ for the same
lattice strip with $L_y=3$ \cite{tk} but has a larger size.  Since the dominant
$\lambda$'s in these three regions have coefficients with order $O(1)$,
$O(q)$, and $O(q^2)$, which are called level 0, 1, and 2 in
\cite{lse0004}, naively one may infer this to be true for arbitrary $L_y$. 
The locus ${\cal B}$ determined from the eight general $\lambda$'s given
in \cite{lse0004} for $3 \le L_y \le 10$ is shown in Fig. \ref{sqkpxpy},
where the envelope of the outer boundary of ${\cal B}$ does increase in
size as $L_y$ increases. However, for $L_y=8, 9$, and $10$, an extra small
regions $R_4$ appears around $q=7, 8$, and $8$, respectively.  Therefore,
even for such simple structure of $\lambda$'s (in the form of polynomial),
the locus ${\cal B}$ may not remain simple for sufficiently large $L_y$. 

\vspace{10mm}

Acknowledgment: I would like to thank Prof. R. Shrock for helpful
discussions.

\vfill
\eject

\end{document}